\documentclass[fleqn,usenatbib]{mnras}

\usepackage[flushleft]{threeparttable}
\usepackage{pifont}
\usepackage{longtable}
\usepackage[T1]{fontenc}
\usepackage{amsmath} % or simply amstext
\newcommand{\angstrom}{\textup{\AA}}

\DeclareRobustCommand{\VAN}[3]{#2}
\let\VANthebibliography\thebibliography
\def\thebibliography{\DeclareRobustCommand{\VAN}[3]{##3}\VANthebibliography}
\usepackage{graphicx}	% Including figure files
\usepackage{amsmath}	% Advanced maths commands
\usepackage{amssymb}	% Extra maths symbols
\usepackage{newtxtext,newtxmath}
\def\gax{\mathrel{\raise.3ex\hbox{$>$}\mkern-14mu\lower0.6ex\hbox{$\sim$}}}
\def\lax{\mathrel{\raise.3ex\hbox{$<$}\mkern-14mu\lower0.6ex\hbox{$\sim$}}}
\def\gtorder{\mathrel{\raise.3ex\hbox{$>$}\mkern-14mu
             \lower0.6ex\hbox{$\sim$}}}
\def\ltorder{\mathrel{\raise.3ex\hbox{$<$}\mkern-14mu
             \lower0.6ex\hbox{$\sim$}}}

\title[An interacting red giant binary]{The ``Giraffe'': Discovery of a stripped red giant in an interacting binary with a ${\sim}2~M_\odot$ lower giant}
\author[T. Jayasinghe et al.]{T. Jayasinghe$^{1,2}$\thanks{E-mail: jayasinghearachchilage.1@osu.edu},
Todd A. Thompson$^{1,2}$,
C. S. Kochanek$^{1,2}$,
K. Z. Stanek$^{1,2}$,
D. M. Rowan$^{1,2}$,
\newauthor 
D. V. Martin$^{1,3,2}$,
Kareem El-Badry$^{4,5,6}$,
P. J. Vallely$^{1,2}$,
J. T. Hinkle$^{7}$,
D. Huber$^{7}$,
H. Isaacson$^{8,9}$,
\newauthor 
J. Tayar$^{7,10}$,
K. Auchettl$^{11,12,13}$,
I. Ilyin$^{14}$,
A. W. Howard$^{15}$,
C. Badenes$^{16}$
\\
$^{1}$Department of Astronomy, The Ohio State University, 140 West 18th Avenue, Columbus, OH 43210, USA\\
$^{2}$Center for Cosmology and Astroparticle Physics, The Ohio State University, 191 W. Woodruff Avenue, Columbus, OH 43210, USA\\
$^{3}$NASA Sagan Fellow\\
$^{4}$Center for Astrophysics $|$ Harvard \& Smithsonian, 60 Garden Street, Cambridge, MA 02138, USA\\
$^{5}$Harvard Society of Fellows, 78 Mount Auburn Street, Cambridge, MA 02138\\
$^{6}$Max-Planck Institute for Astronomy, K\"onigstuhl 17, D-69117 Heidelberg, Germany\\
$^{7}$Institute for Astronomy, University of Hawai`i, 2680 Woodlawn Drive, Honolulu, HI 96822, USA\\
$^{8}$Department of Astronomy,  University of California Berkeley, Berkeley CA 94720, USA\\
$^{9}$Centre for Astrophysics, University of Southern Queensland, Toowoomba, QLD, Australia\\
$^{10}$Department of Astronomy, University of Florida, Bryant Space Science Center, Stadium Road, Gainesville, FL 32611, USA \\
$^{11}$OzGrav, School of Physics, The University of Melbourne, Parkville, Victoria 3010, Australia\\
$^{12}$ARC Centre of Excellence for All Sky Astrophysics in 3 Dimensions (ASTRO 3D)\\
$^{13}$Department of Astronomy and Astrophysics, University of California, Santa Cruz, CA 95064, USA\\
$^{14}$Leibniz Institute for Astrophysics Potsdam (AIP), An der Sternwarte 16, D-14482 Potsdam, Germany\\
$^{15}$Department of Astronomy, California Institute of Technology, Pasadena, CA 91125, USA\\
$^{16}$Department of Physics and Astronomy and Pittsburgh Particle Physics, Astrophysics and Cosmology Center (PITT PACC),\\
University of Pittsburgh, 3941 O‘Hara Street, Pittsburgh, PA 15260, USA\\
}

\date{Accepted XXX. Received YYY; in original form ZZZ}

\pubyear{2022}

\begin{document}
\label{firstpage}
\maketitle

% Abstract of the paper
\begin{abstract}
We report the discovery of a stripped giant + lower giant binary, 2M04123153+6738486 (2M0412), identified during a search for non-interacting compact object---star binaries. 2M0412 is an evolved ($T_{\rm eff, giant}\simeq4000$~K), luminous ($L_{\rm giant}\simeq150~L_\odot$) red giant in a circular $P=81.2$~d binary. 2M0412 is a known variable star previously classified as a semi-regular variable. The cross-correlation functions of follow-up Keck/HIRES and LBT/PEPSI spectra show an RV-variable second component with implied mass ratio $q=M_{\rm giant}/M_{\rm comp}\simeq0.20\pm0.01$. The ASAS-SN, ATLAS, \textit{TESS} and ZTF light curves show that the giant is a Roche lobe filling ellipsoidal variable with an inclination of $49.4^\circ{}\pm{0.3^{\circ}}$, and a giant mass of $M_{\rm giant}=0.38\pm0.01~ M_\odot$ for a distance of $\simeq3.7$~kpc. The mass of the giant indicates that its envelope has been stripped. The giant companion on the lower red giant branch has a mass of $M_{\rm comp}=1.91\pm0.03~M_\odot$  with $T_{\rm eff, comp}\simeq5000$~K, $L_{\rm comp}\simeq60~L_\odot$ and $R_{\rm comp}\simeq11~R_\odot$. We also identify an orbital phase dependent, broad $\rm H\alpha$ emission line which could indicate ongoing accretion from the stripped red giant onto the companion. 
\end{abstract}

% Select between one and six entries from the list of approved keywords.
% Don't make up new ones.
%\clearpage

%%%%%%%%%%%%%%%%%%%%%%%%%%%%%%%%%%%%%%%%%%%%%%%%%%

%%%%%%%%%%%%%%%%% BODY OF PAPER %%%%%%%%%%%%%%%%%%

\section{Introduction}

The mass distribution of neutron stars and stellar mass black holes is closely tied to the evolution of massive stars and supernovae (e.g., \citealt{Sukhbold2016, Pejcha2015, Woosley2020}). The relationship between the pre-supernova mass of the massive star and the type of the compact remnant left behind is complicated and depends on the chemical composition of the star, mass-loss rates, supernova explosion physics and binary interactions (e.g., \citealt{Sukhbold2016, Patton2021}). While the fates of these massive stars are hard to predict, dead massive stars have left behind a plethora of compact remnants in our Galaxy -- there are predict to be about ${\sim}10^8$ stellar mass BHs and ${\sim}10^9$ neutron stars in the Milky Way \citep{Brown1994}.  

%For example, massive stars that explode as Type IIp supernovae and form neutron stars have masses in the range ${\sim}8.4-18.6~M_\odot$ (e.g., \citealt{Kochanek2020}). However, the population of massive stars that produce BHs is largely unconstrained except for the ${\sim}20-25~M_\odot$ progenitor of the failed supernovae candidate NGC6946-BH1 (e.g., \citealt{Adams2017,Basinger2021}). This complicates our understanding of the predicted BH mass function at birth \citep{Woosley2020}. 

A well characterized sample of neutron stars and black holes is necessary to better understand massive stars. However, this is a challenging task because the vast majority of compact objects are electromagnetically dark. To date, most mass measurements for neutron stars and black holes come from pulsar and accreting binary systems selected from radio, X-ray, and gamma-ray surveys (see, for e.g., \citealt{Champion2008,Liu2006,Ozel2010,Farr2011}), and from the LIGO/Virgo detections of merging systems (see, for e.g., \citealt{TheLIGOScientificCollaboration2021, Abbott2016,Abbott2017NS}). These interacting systems are, however, a small minority of compact object binaries, and the far larger population of non-interacting systems is essentially unexplored. While non-interacting binaries are harder to find, they must be discovered and characterized in order to fully understand the numbers, properties, formation mechanism and evolutionary paths of the interacting systems. 

%The populations of BHs observed as X-ray binaries and mergers are both heavily biased samples. For X-ray binaries, the companion must either fill its Roche lobe or have a modest separation and a strong wind. Compact objects discovered through gravitational wave observations come from the very small fraction of surviving binaries that are on very close orbits leading to a merger. These interacting systems are, however, a small minority of compact object binaries, and the far larger population of non-interacting systems is essentially unexplored. While non-interacting binaries are harder to find, they must be discovered and characterized in order to fully understand the numbers, properties, formation mechanism and evolutionary paths of the interacting systems. 

Rapid advances in time-domain astronomy and precision \textit{Gaia} astrometry \citep{Lindegren2021} provide promising pathways for future discoveries of non-interacting compact objects. For example, \citet{Chawla2021} estimated that ${\sim}30-300$ non-interacting black holes are detectable in binaries around luminous companions using astrometry from \textit{Gaia}. Similarly, \citet{Shao2019} used binary population synthesis models to estimate that there are ${\sim}10^3$ detached non-interacting black holes in the Milky Way, with $10^2$ of these systems having luminous companions that are brighter than $G{\sim}20$ mag. In addition to astrometry, targeted searches combining high-cadence photometry and sparsely sampled radial velocities from wide-field time-domain surveys are a promising method to discover more systems (e.g., \citealt{Thompson2019,Zheng2019,Rowan2021,Trimble1969}). 

Only a handful of convincing non-interacting compact objects have been discovered thus far. Three non-interacting BH candidates have been discovered in globular clusters: one by \citet{Giesers2018} in NGC 3201 (minimum black hole mass  $M_{\rm BH} = 4.36 \pm0.41$\,M$_\odot$), and two by \citet{Giesers2019} in NGC 3201 ($M_{\rm BH}\sin(i) = 7.68 \pm0.50$\,M$_\odot$ and $M_{\rm BH}\sin(i) = 4.4 \pm2.8$\,M$_\odot$). These globular cluster systems, if they indeed contain black holes, likely have formation mechanisms that are very different from those of field black hole binaries because the high stellar densities allow formation mechanisms which do not operate for field stars. A single convincing non-interacting BH candidate has been found in the field. \citet{Thompson2019} discovered a low-mass ($M_{\rm BH}\simeq3.3_{-0.7}^{+2.8}~M_\odot$) non-interacting black hole candidate in the field on a circular orbit at $\rm P_{\rm orb}\sim83\,d$ around a spotted giant star. 

However, the search for non-interacting compact objects is challenging, and numerous false positives have been identified. The binary LB-1 was initially thought to host an massive stellar black hole ($M_{\rm BH}\simeq68_{-3}^{+11}~M_\odot$, \citealt{Liu2019}), but was later found to have a much less massive companion that was not necessarily a compact object (see, for e.g., \citealt{Shenar2020,Irrgang2020,Abdul-Masih2020,El-Badry2020a}). The naked-eye system HR 6819 was claimed to be a triple system with a detached black hole with $M_{\rm BH} = 6.3 \pm0.7~M_\odot$ \citep{Rivinius2020}, but was later found to be a binary system with a rapidly rotating Be star and a slowly rotating B star \citep{El-Badry2020b,Bodensteiner2020}. Recently, NGC 1850 BH1 was claimed to be a binary displaying ellipsoidal variability in the LMC with $M_{\rm BH} = 11.1 _{-2.4}^{+2.1}~M_\odot$ \citep{Saracino2021}, but was later argued to be a stripped B star binary \citep{El-Badry2021}. Another example of a BH imposter was the system NGC 2004 \#115, claimed to be a triple system consisting of a Be star on a tertiary orbit and an inner binary of a B star and a ${\simeq}25~M_\odot$ black hole \citep{Lennon2021}. \citet{El-Badry2021NGC2004} later argued that the orbital inclination was underestimated by assuming tidal synchronization, and that the companion to the B star was more likely a ${\sim}2-3~M_\odot$ main sequence star. \citet{Jayasinghe2021} identified the nearby, nearly edge-on $\rm P_{\rm orb}=59.9$~d circular binary V723~Mon as a candidate for a compact object---star binary. \citet{ElBadry2022} later showed that V723 Mon is better explained by a stripped red giant in a binary around a massive (${\sim}2.8M_\odot$), rapidly rotating subgiant. A common theme to these cases is an overestimate of the mass of the observed star based on its luminosity and the assumption of single star evolution for a binary where mass transfer has greatly reduced the mass of the more luminous star. 

2M04123153+6738486 (hereafter 2M0412) is a luminous red-giant in the Camelopardalis constellation with J2000 coordinates $(\alpha,\delta)=(63.13141144391^\circ,67.64683129470^\circ)$. It was first classified as an `NSINE' variable star (ATO J063.1314+67.6468) in the the Asteroid Terrestrial-impact Last Alert System (ATLAS; \citealt{Tonry2018,Heinze2018}) catalog of variable stars with a period of ${\sim}80.36$~days. It was classified as a semi-regular variable (ASASSN-V J041231.49+673848.6, ZTF J041231.52+673848.6) by the All-Sky Automated Survey for SuperNovae (ASAS-SN; \citealt{Jayasinghe2018, Jayasinghe2021var,Shappee2014}), and the Zwicky Transient Facility (ZTF; \citealt{Chen2020, Bellm2019}) with periods of ${\sim}40.65$~days and  ${\sim}41.20$~days respectively. 

2M0412 has 4 radial velocity measurements in the Apache Point Observatory Galactic Evolution Experiment DR16 (APOGEE; \citealt{Gunn2006,Blanton2017,Wilson2019}) with a maximum velocity difference of $\Delta\rm RV_{\rm max}\sim76$~km/s and a maximum observed acceleration between epochs of ${\sim}1.7$~km/s/day \citep{Thompson2019}. This led us to re-examine the light curves to realize that 2M0412 was in fact an ellipsoidal variable with an ${\sim}81$~day period. The radial velocity data and the orbital period from the light curves implied a mass function of $f(M)\simeq0.5~M_\odot$, which led us to investigate the system in detail.

Here we discuss our discovery of a ${\sim}1.9~M_\odot$ lower giant companion to the red giant in 2M0412. We initially considered this system as a viable candidate for a compact object--star binary, but a detailed look at the observational evidence suggests that 2M0412 is best described as an interacting binary composed of a stripped red giant and a more massive lower giant companion. We describe the archival data and new observations used in our analysis in Section \ref{section:obssec}. In Section \ref{section:results}, we analyze the photometric and spectroscopic observations to derive the parameters of the binary system and the red giant. In Section \ref{section:bpass}, we compare the observations to binary population synthesis models. In Section \ref{section:darkcomp}, we discuss the nature of the companion. We present a summary of our results in Section \ref{section:summary}. 

\section{Observations}
\label{section:obssec}
Here we present observations, both archival and newly acquired, that are used in our analysis of 2M0412.

\subsection{Distance, Kinematics and Extinction}

In \textit{Gaia} EDR3 \citep{GaiaEDR3MAIN}, 2M0412 is {\verb"source_id="490450875403620352}. Its raw EDR3 parallax of $\varpi_{\rm EDR3}=0.2310 \pm0.0130 \,\rm mas$ implies a distance of $d=1/\varpi=4329\pm 245$~pc. We correct the reported EDR3 parallax using the zero-point correction (${\simeq39.89~\mu\rm as}$) described in \citet{Lindegren2021} which is dependent on the color, magnitude, and sky position\footnote{We calculate the parallax zero-point using the python script provided in \url{https://gitlab.com/icc-ub/public/gaiadr3_zeropoint}}. With this correction, we obtain $\varpi_{\rm EDR3, L21}=0.2709 \pm0.0130 \,\rm mas$ ($d=1/\varpi=3691\pm 178$~pc). The global zero-point parallax offset in EDR3 is $-17~\mu\rm as$ \citep{Lindegren2021astro}, and if we instead use this correction, we obtain $d=1/\varpi=4032\pm 212$~pc. The zero-point correction from \citet{Lindegren2021} differs from the global zero-point correction by ${\sim}-23~\mu\rm as$. The probabilistic photogeometric distance estimate from \citet{Bailer-Jones2021} is $d=3743 ^{+133} _{-161}\,\rm pc$. The spectrophotometric parallax estimate from \citet{Hogg2019} is $\varpi_{\rm spec}=0.2284 \pm0.0106 \,\rm mas$ ($d=1/\varpi=4378\pm 204$~pc). The astrometric solution does not show significant excess noise and its renormalized unit weight error (RUWE) of $1.046$ is not indicative of problems in the parallax or strong resolved binary motion. We adopt a distance of $d=3.7$~kpc as our default.

At this distance, 2M0412 is ${\sim}830$~pc above the midplane. Its proper motion in EDR3 is $\mu_{\alpha}=-0.346\pm0.008 \, \rm mas\;yr^{-1}$, and $\mu_{\delta}=0.686\pm0.011 \, \rm mas\;yr^{-1}$.  Combining this with the systemic radial velocity from $\S$\ref{section:orbit}, the definition of the local standard of rest (LSR) from \citet{Schonrich2010}, and using \verb"BANYAN" \citep{Gagne2018} for the calculations, the 3D space motion of 2M0412 relative to the LSR is $(U,V,W)_{\rm LSR}=(45.2,-19.5,-5.1)$ $\rm km~s^{-1}$. We calculated the thin disk, thick disk and halo membership probabilities based on these velocities following \citet{Ramirez2013} to obtain $P(\rm thin)\simeq98.5\%$, $P(\rm thick)\simeq1.5\%$ and $P(\rm halo)\simeq0\%$, respectively. This suggests that this system is a kinematically normal object in the thin disk.

\subsection{Light Curves}

We analyzed well-sampled light curves from ASAS-SN, ATLAS and ZTF, along with a densely sampled but phase-incomplete light curve from the Transiting Exoplanet Survey Telescope (TESS).

The ASAS-SN \citep{Shappee2014,Kochanek2017} survey obtained $V$-band and $g$-band light curves of 2M0412 spanning from December 2014 to November 2021 (${\sim}31$ orbits). 2M0412 clearly varies in the ASAS-SN light curves, with two equal maxima and two minima when phased with the orbital period.
We determined the photometric period using \verb"Period"04 \citep{Lenz2005}. The dominant ASAS-SN periods of $P_{\rm ASAS-SN,V}=40.5930\pm0.0572$~d and $P_{\rm ASAS-SN,g}=40.6090\pm 0.08442$~d correspond to $P_{\rm orb}/2$ for ellipsoidal variability. We find an orbital period of
\begin{equation}
     P_{\rm orb,ASAS-SN~V}=81.1861\pm0.1144\,{\rm d}.
	\label{eq:period}
\end{equation} 

We retrieved $g$ and $r-$band photometry from the ZTF DR7 catalog \citep{Masci2019} spanning April 2018 to April 2021 (${\sim}13$ orbits). The dominant periods in the ZTF data, $P_{\rm ZTF,g}=40.7195\pm0.0770$~d, $P_{\rm ZTF,r}=40.5224\pm0.0753$~d, again correspond to $P_{\rm orb}/2$. We also retrieved photometry from the ATLAS survey \citep{Tonry2018} in the ATLAS `$c$' (cyan) and `$o$' (orange) filters. The ATLAS observations span from August 2017 to March 2021 (${\sim}16$ orbits). We obtain photometric periods of $P_{\rm ATLAS, c}=40.5497\pm0.0633$~d, $P_{\rm ATLAS, o}=40.5447\pm0.0627$~d. The differences between the ASAS-SN, ZTF and ATLAS photometric period estimates are not statistically significant.

2M0412 (\verb"TIC" 102780470) was observed by \textit{TESS} \citep{Ricker2015} in Sector 19, and the 27 days of observations span [0.22,0.52] in orbital phase where the phase of the RV maximum is $0.75$. We analyzed the \textit{TESS} data using the adaptation of the ASAS-SN image subtraction pipeline for analyzing \textit{TESS} full-frame images (FFIs) described in \cite{Vallely2020}. The light curve does not include epochs where the observations were compromised by the spacecraft's momentum dump maneuvers.

\begin{figure*}
	
	\includegraphics[width=0.7\textwidth]{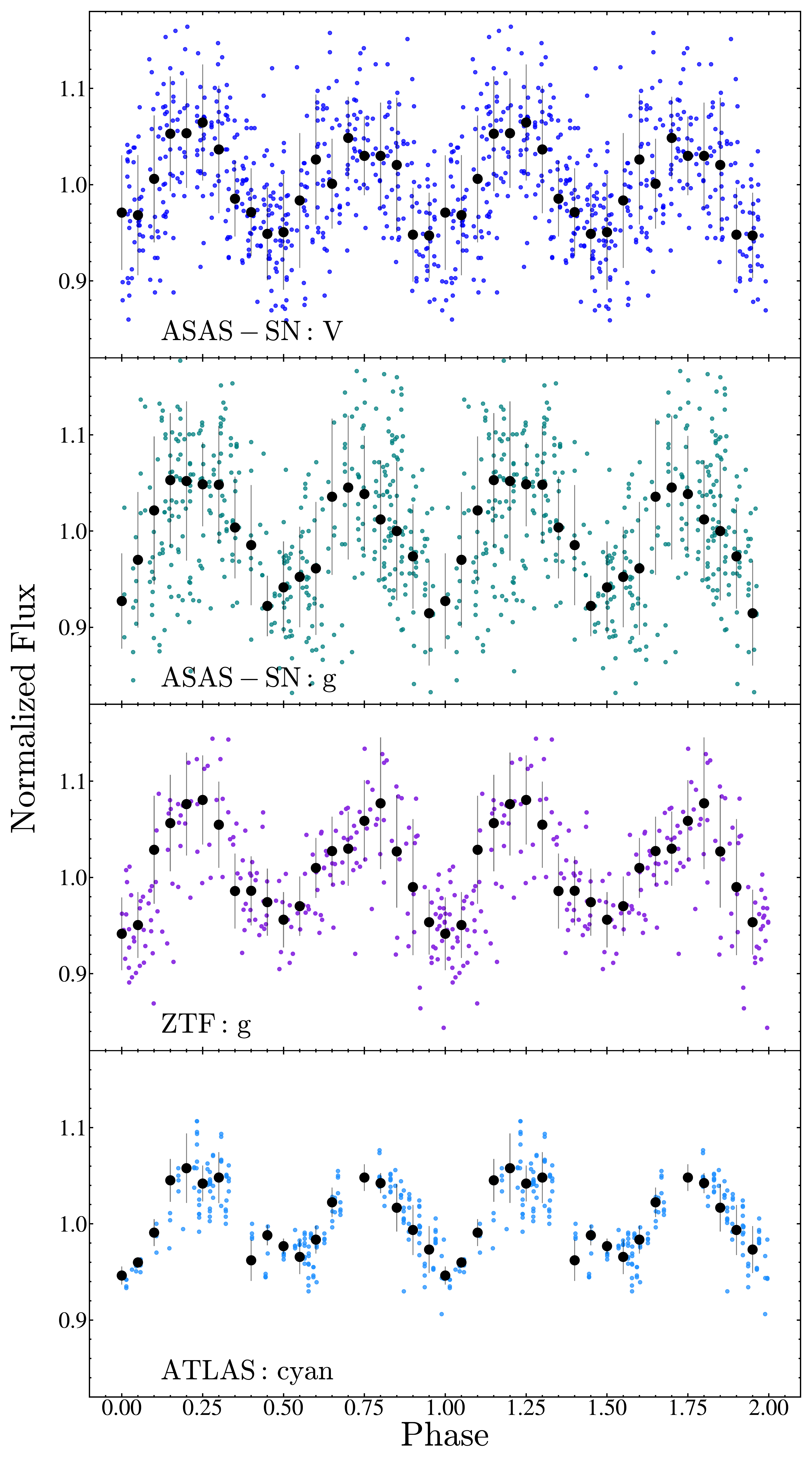}
		\vspace{-0.2cm}	
    \caption{The normalized ASAS-SN $V$, ASAS-SN $g$, ZTF $g$ and ATLAS $c$-band light curves for 2M0412 as a function of orbital phase (defined with the epoch of maximum RV at $\phi=0.75$). The binned light curve for each data set is shown in black. The error bars for the binned light curve show the error on the mean.}
    \label{lcsblue}
\end{figure*}

\begin{figure*}
	
	\includegraphics[width=0.7\textwidth]{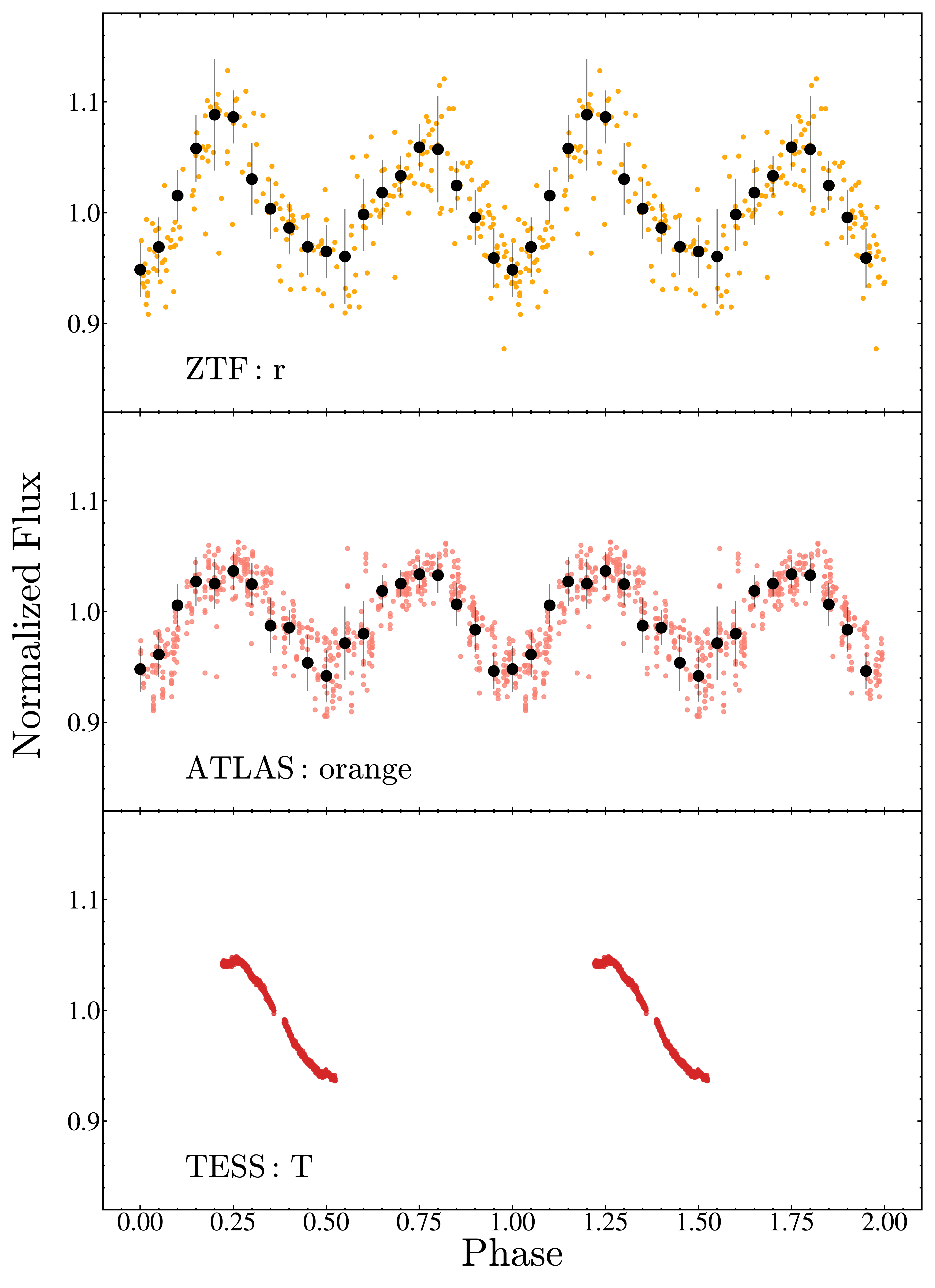}
		\vspace{-0.2cm}	
    \caption{The normalized ZTF $r$, ATLAS $o$ and \textit{TESS} $T$-band light curves for 2M0412. The format is the same as Figure \ref{lcsblue}.}
    \label{lcsred}
\end{figure*}

Figures \ref{lcsblue} and \ref{lcsred} present the ASAS-SN, ATLAS, ZTF and \textit{TESS} light curves for 2M0412 phased at the orbital period of $P_{\rm orb}\simeq81.2$~d. When phased with the orbital period, the light curves have two equal maxima and two unequal minima, which is typical of ellipsoidal variability. The light curves in the bluer filters tend to have more scatter than those in the redder filters. This is likely physical in nature and could indicate the presence of spots on the giant.

\subsection{UV photometry}

We obtained Swift UVOT \citep{roming05} images in the $UVM2$ (2246 \AA) band \citep{poole08} through the Target of Opportunity (ToO) program (Target ID number 14417). We only obtained images in the $UVM2$ band because the \textit{Swift} $UVW1$ and $UVW2$ filters have significant red leaks that make them unusable in the presence of the emission from the cool giant. Each epoch of UVOT data includes multiple observations, which we combined using the \texttt{uvotimsum} package. We then used \texttt{uvotsource} to extract source counts using a 5\farcs{0} radius aperture centered on the star. We measured the background counts using a source-free region with a radius of 40\farcs{0} and used the most recent calibrations \citep{poole08, breeveld10} and taking into account the recent update to the sensitivity corrections for the \textit{Swift} UV filters\footnote{\url{https://www.swift.ac.uk/analysis/uvot/index.php}}. We were unable to detect 2M0412 in the \textit{Swift} $UVM2$ data. The $UVM2$ observations are summarized in Table \ref{tab:swift}.

2M0412 was previously observed as part of the \textit{GALEX} Medium Imaging survey (MIS) and All-Sky Imaging Survey (AIS) in January 2011 and January 2004, respectively \citep{Morrissey2007}. As part of the MIS survey, 2M0412 was observed for 1689 seconds in the NUV filter. Additionally, 2M0412 was observed for 232 seconds in both the \textit{NUV} and \textit{FUV} filters. We do not detect 2M0412 in any of the \textit{GALEX} images. In the \textit{NUV}, we adopt the $22.7$~AB mag $5\sigma$ upper limit from the longer MIS exposure \citep{Bianchi2014}. In the \textit{FUV}, we adopt the $19.9$~AB mag $5\sigma$ upper limit from the AIS exposure. The \textit{$GALEX$} observations are summarized in Table \ref{tab:swift}.

\begin{table*}
	\centering
	\caption{\textit{GALEX} and \textit{Swift} UV observations}
	\label{tab:swift}
\begin{tabular}{rrrrrrr}
		\hline
		 JD & Date & Phase & Filter & AB mag limit & $\sigma$\\
		\hline
2459453.807 & 2021-08-27 & 0.076 & \textit{Swift} \textit{UVM2} & $>22.78$ & $3\sigma$\\
2459488.654 & 2021-10-01 & 0.506 & \textit{Swift} \textit{UVM2} & $>22.67$ & $3\sigma$\\
2455572.568 & 2011-01-11 & 0.259 & \textit{GALEX} \textit{NUV} (MIS) & $>22.7$ & $5\sigma$\\
2453017.625 & 2004-01-13 & 0.782 & \textit{GALEX} \textit{FUV} (AIS) & $>19.9$ & $5\sigma$\\

\hline
\end{tabular}
\end{table*}

\subsection{Spectroscopic observations}

To better sample the radial velocities, we obtained 10 additional spectra. These observations, along with the archival APOGEE DR16 observations, are summarized in Table \ref{tab:spec}. 
Using the HIRES instrument \citep{Vogt1994} on Keck I, we obtained 8 spectra with $R \approx 60000$ between Aug 14 2021 and Nov 19 2021 using the standard California Planet Search (CPS) setup \citep{Howard2010}. The exposure times ranged from 292 to 515 seconds.
We also obtained very high resolution ($R \approx 130,000$) spectra on 10 Sep 2021 and 02 Oct 2021 using the Potsdam Echelle Polarimetric and Spectroscopic Instrument (PEPSI; \citealt{Strassmeier2015}) on the Large Binocular Telescope. We used the $200~\mu$m fiber and 2 cross-dispersers (CD). The data were processed as described in \citet{Strassmeier2018}. The total integration time for each epoch was 70 minutes, and the 2 CDs cover the wavelength ranges $4758-5416~{\angstrom}$ and $6244-7427~{\angstrom}$. The HIRES, PEPSI and MODS observations are summarized in Table \ref{tab:spec}.

\begin{table*}
	\centering
	\caption{Radial velocities from APOGEE DR16, HIRES, and PEPSI.}
	\label{tab:spec}
\begin{tabular}{rrrrrr}
		\hline
		 BJD & Date & Phase & RV& $\sigma_{RV}$ & Instrument\\
		 (TDB) &  &  & ($\rm km s^{-1}$) & ($\rm km s^{-1}$) & \\		 
		\hline
2457279.99438 & 2015-09-14 & 0.295 & $-$86.888 & 0.031 & APOGEE DR16 \\
2457292.96813 & 2015-09-27 & 0.455 & $-$59.246 & 0.016 & APOGEE DR16 \\
2457320.92338 & 2015-10-25 & 0.799 & $-$7.964 & 0.023 & APOGEE DR16 \\
2457323.87005 & 2015-10-28 & 0.835 & $-$11.194 & 0.017 & APOGEE DR16 \\
2459441.12722 & 2021-08-14 & 0.920 & $-$26.94 & 0.10 & HIRES \\
2459442.11471 & 2021-08-15 & 0.932 & $-$30.43 & 0.10 & HIRES \\
2459457.11748 & 2021-08-30 & 0.117 & $-$74.32 & 0.10 & HIRES \\
2459467.98696 & 2021-09-10 & 0.251 & $-$87.423 & 0.060 & PEPSI \\
2459472.12548 & 2021-09-14 & 0.302 & $-$86.11 & 0.10 & HIRES \\
2459485.12694 & 2021-09-27 & 0.462 & $-$57.29 & 0.10 & HIRES \\
2459489.96760 & 2021-10-02 & 0.522 & $-$41.928 & 0.034 & PEPSI \\
2459502.97527 & 2021-10-15 & 0.682 & $-$9.84 & 0.10 & HIRES \\
2459511.98487 & 2021-10-24 & 0.793 & $-$7.61 & 0.10 & HIRES \\
2459537.92734 & 2021-11-19 & 0.113 & $-$73.81 & 0.10 & HIRES \\

\hline
\end{tabular}
\end{table*}

\subsection{X-ray data}

We analyzed the X-ray observations from the \textit{Swift} X-Ray Telescope (XRT; \citealt{Burrows2005}) taken simultaneously with the $UVM2$ observations. Individual exposure times ranged from 120 to 1650 seconds, for a total of 4010 seconds. All XRT observations were reprocessed using the \textit{Swift} \textsc{xrtpipeline} version 0.13.2 and standard filter and screening criteria\footnote{\url{http://swift.gsfc.nasa.gov/analysis/xrt_swguide_v1_2.pdf}} and the most up-to-date calibration files. To increase the signal to noise of our observations, we combined all cleaned individual XRT observations using \textsc{XSELECT} version 2.4g. To place constraints on the presence of X-ray emission (see $\S$\ref{section:xrays}), we used a source region with a radius of 30 arcsec centered on the position of 2M0412 and a source-free background region with a radius of 150 arcsec located at RA = 04:12:38.10, Dec = 67:43:00.86 (J2000).

\section{Results}
\label{section:results}

Here we present our analyses of the observations described in $\S$\ref{section:obssec}. In $\S$\ref{section:orbit}, we fit Keplerian models to the radial velocities and derive a SB2 spectroscopic orbit for 2M0412. In $\S$\ref{section:giant}, we characterize the red giant using its spectral energy distribution (SED) and spectra. In $\S$\ref{section:phoebe}, we model the ellipsoidal variations of the red giant using multi-band light curves and the binary modeling tool \verb"PHOEBE" to derive the masses of the red giant and the companion. In $\S$\ref{section:balmer}, we discuss the broad Balmer $\rm H\alpha$ emission and its orbital phase dependent variability. In $\S$\ref{section:xrays}, we discuss the X-ray observations.

\subsection{Keplerian orbit models}\label{section:orbit}

\begin{figure*}
	\includegraphics[width=0.65\textwidth]{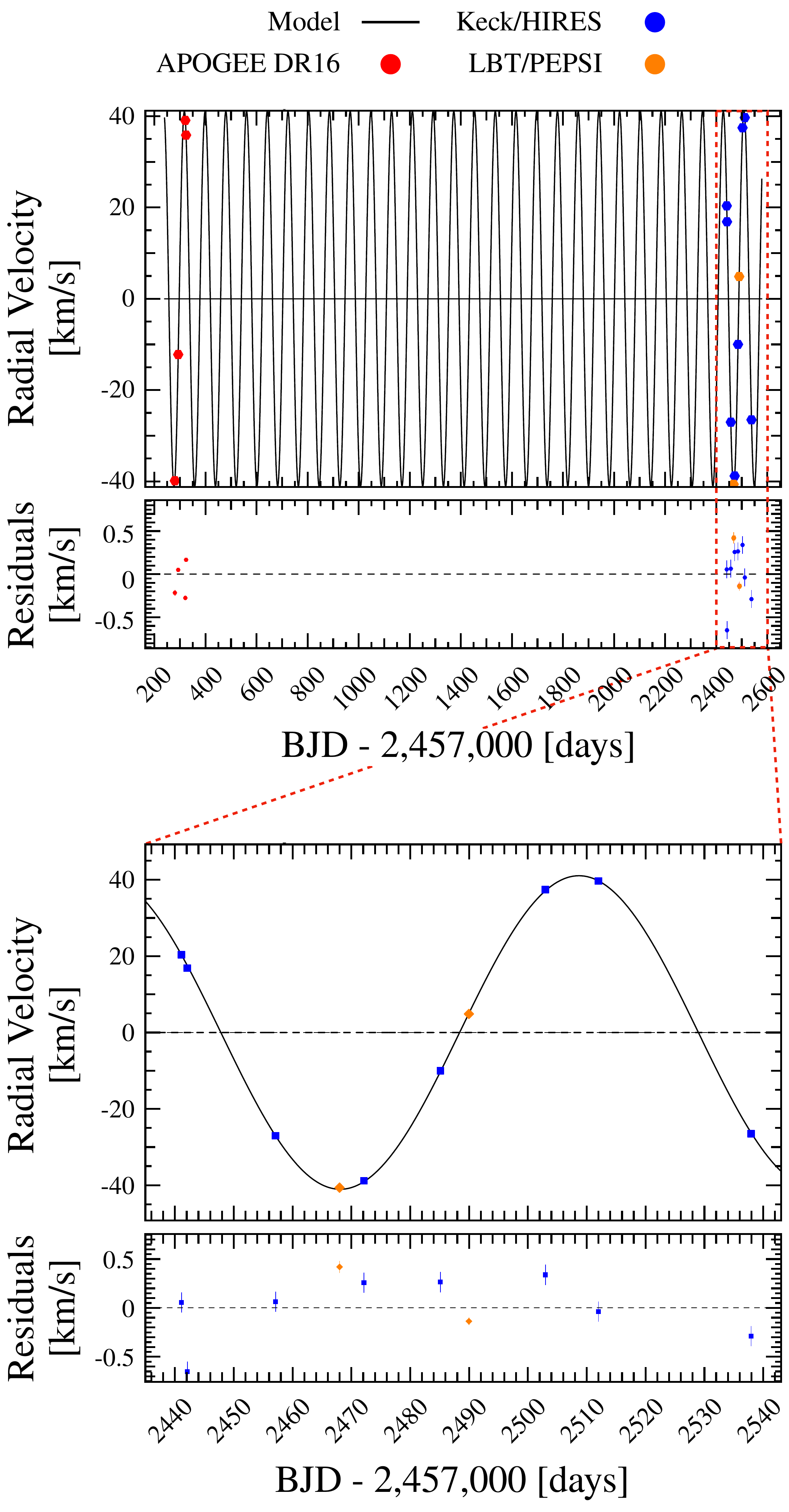}
    \caption{\textsc{Yorbit} radial velocity fit to the APOGEE, HIRES and PEPSI data. The top figure shows the entire observational timespan. The bottom figure is zoomed into our follow-up measurements with HIRES and PEPSI, covering slightly more than one orbital period.}
    \label{yorbitRVs}
\end{figure*}

We fit Keplerian orbits to the APOGEE DR16, HIRES and PEPSI radial velocities using two separate codes. First, we use \verb"TheJoker" \citep{Price-Whelan2017} Monte Carlo sampler. We first fit the joint data set using a broad uniform prior in period spanning [1~d,1000~d]. We obtain a unimodal solution at $P_{\rm orb}\simeq82.6$~d, which is close to the orbital period from the light curves. We then reran \verb"TheJoker" with a Gaussian prior on the period of $P_{\rm orb}=81.2\pm0.5$~d. In these joint fits we also include additional parameters to allow for any velocity zero point offsets between the three instruments and set the argument of periastron to $\omega=0$ after confirming that there was no evidence for a non-zero ellipticity. The results of the fits are summarized in Table \ref{tab:orbit}. The mass functions are well-constrained and mutually consistent, and the elliptical models yield a very small, non-zero ellipticity that is not significant. As a second, independent fit, we use the \textsc{Yorbit} genetic algorithm \citep{Segransan2011}, frequently used in the discovery of extra-solar planets (e.g. \citealt{Mortier2016,Martin2019}). We find orbital parameters consistent with \verb"TheJoker" results. In Fig.~\ref{yorbitRVs} we plot the \textsc{Yorbit} fit to the data. Given that the orbit is circular, we define the phase relative to the epoch of maximal radial velocity $T_{\rm RV, max}=2457722.795\pm0.130\,{\rm d}$ instead of the epoch of periastron. We define orbital phases so that $T_{\rm RV, max}$ corresponds to $\phi=0.75$, the companion and the giant would be eclipsed at $\phi=0.5$ and $\phi=0$, respectively, for an edge-on orbit. 

%%While we fit the RVs independently of the photometry here, for the remainder of this paper we use the orbit derived from the combined RV+light curve fits in $\S$\ref{section:phoebe}. 

\begin{table*}
	\centering
	\caption{Orbital Elements for 2M0412}
	\label{tab:orbit}
\begin{tabular}{rrrrr}
		\hline
		 Parameter & Joint RV data+ \verb"TheJoker" & Joint RV data+ \verb"TheJoker" & HIRES+ \verb"TheJoker" &  Joint RV data + \verb"yorbit" \\
		\hline
	$P_{\rm orb} (\rm d)$ & $81.439\pm0.001$ & $81.120\pm0.001$ & $81.136\pm0.103$ & $81.1643\pm0.0029$ \\
	$K~(\rm km ~s^{-1})$ & $41.041\pm0.138$ &  $40.976\pm0.144$ & $41.016\pm0.138$ & $41.038\pm0.138$ \\	
	$e$ & -- & $0.004\pm0.003$ & -- & -- \\
	$\omega~(^\circ)$ & 0 (fixed) & $341\pm71$ & 0 (fixed) & 0 (fixed) \\	
	${v_\gamma}~(\rm km~ s^{-1})$  &  $-47.137\pm0.108$ &  $-47.096\pm0.173$ & $-47.217\pm0.098$ & $-47.031\pm0.082$ \\
	$a_{\rm giant}\sin i$ $(R_{\odot})$  & $65.846\pm0.214$ & $65.996\pm0.359$ & $65.840\pm0.285$ & $65.840\pm0.285$ \\	
    \hline    
	$f(M)$ $(M_{\odot})$  & $0.583\pm0.006$ & $0.578\pm0.006$ & $0.580\pm0.006$ & $0.581\pm0.002$ \\

\hline
\end{tabular}
\end{table*}

\begin{figure}
	\includegraphics[width=0.5\textwidth]{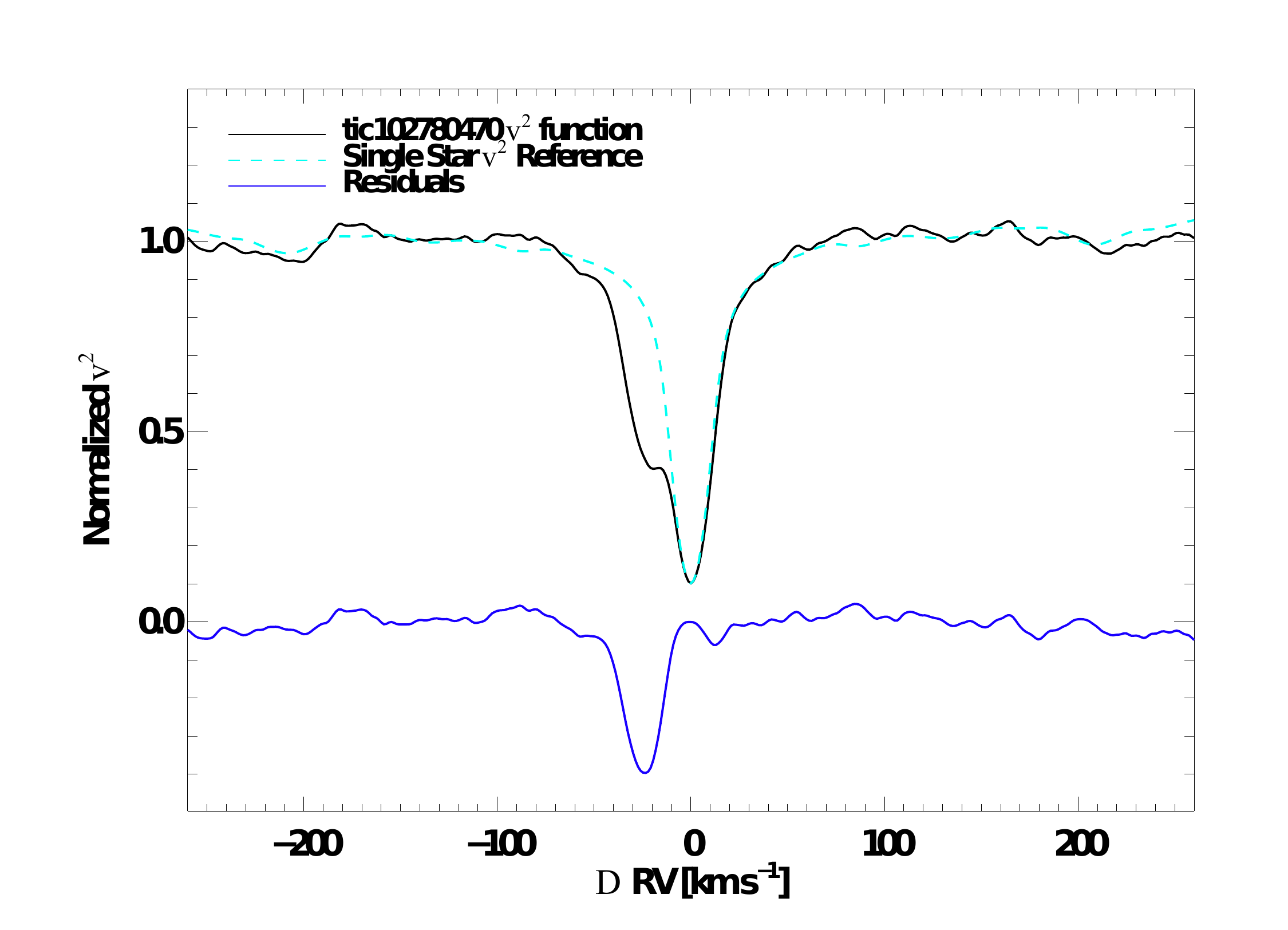}
    \caption{$\chi^2$ as a function of the Doppler shift for the HIRES epoch at phase 0.92. The black line shows the observed reamatch $\chi^2$ function \citep{Kolbl2015} for 2M0412. The light blue dashed line shows the expected $\chi^2$ function for a single star. The dark blue line shows the residual $\chi^2$ function after subtracting the single star reference. }
    \label{rvccf}
\end{figure}

The automated SB2 analysis tool \verb"reamatch" \citep{Kolbl2015} in the Keck CPS pipeline is used to identify double-lined spectroscopic binaries. It finds a secondary peak in the radial velocity cross-correlation function (CCF) for 2M0412 in most of the HIRES spectra (see Figure \ref{rvccf}). \verb"reamatch" only reports an approximate velocity for the second component. To derive more precise estimates, we use \verb"iSpec" to derive RVs for this second component for both the HIRES and PEPSI epochs by cross-correlating the spectra with a synthetic spectrum of a 4250~K red giant. \verb"reamatch" also provides estimates of the secondary temperature and the relative brightness of the secondary in the $V/R$-bands. The secondary temperature is estimated to be  $T_{\rm eff, comp}\sim6000-6100$~K, which lies at the edge of the \verb"reamatch" grid. The relative brightness of the secondary is estimated to range from ${\sim}20\%$ to ${\sim}75\%$ in the $V/R$-bands. 

\begin{figure*}
	\includegraphics[width=0.8\textwidth]{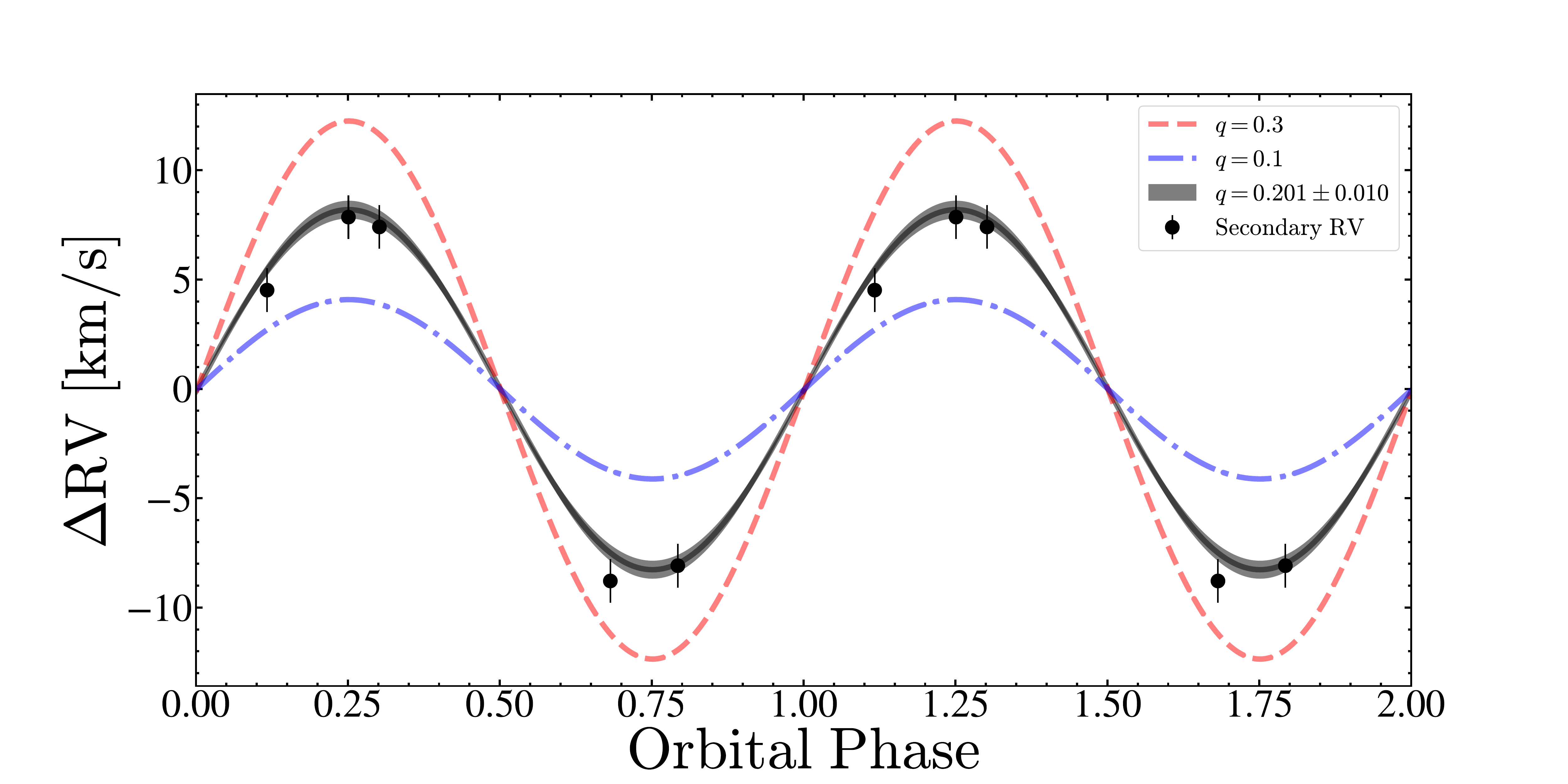}
    \caption{Orbital velocity as a function of phase for the second component in the HIRES and PEPSI spectra. The black shaded region illustrates the expected counter-motion of a component with a mass ratio $q=0.201\pm0.010$. Theoretical RV curves for mass ratios of $q=0.3$ and $q=0.1$ are shown as red dashed and blue dot-dashed lines respectively. }
    \label{rvsb2}
\end{figure*}

Figure \ref{rvsb2} shows the radial velocities for the secondary component after subtracting the systemic velocity $v_\gamma$. If we fix the period and orbital phases based on the orbit of the giant, we find a good fit (see Figure \ref{rvsb2}) to these velocities with $K_{\rm comp}=8.3\pm0.1\, \rm km \,s^{-1}$, which implies $q=M_{\rm giant}/M_{\rm comp}=0.201\pm0.010$ and $a_{\rm comp}\sin i=13.3\pm0.2~R_\odot$. A synchronized, Roche lobe filling star satisfies the relation
\begin{equation}
    \label{eq:qrot}
    v_{\rm rot} \, \textrm{sin}~i/K_{\rm giant}=0.462~q^{1/3}(1+q)^{2/3} 
\end{equation} (see, for e.g., \citealt{Wade1988,Torres2002}). With this relation, we obtain $q=0.21\pm0.03$, which is entirely consistent with the mass ratio derived from the velocities. This indicates that the giant has likely filled its Roche lobe.

Using just the dynamical information ($K_{\rm giant}$, $K_{\rm comp}$ and $P_{\rm orb}$), we get $a_{\rm orb}\sin i=(a_{\rm giant}+a_{\rm comp})\sin i=80.4\pm0.2~R_\odot$. Using Kepler's third law, this corresponds to $M_{\rm tot}\sin^3 i=(M_{\rm giant}+M_{\rm comp})\sin^3 i=1.06\pm0.01~M_\odot$. Using the constraint on $q$, we can derive the minimum mass of the companion and the giant as
\begin{equation}
    M_{\rm comp}\sin^3 i=0.88\pm0.01~M_\odot,
\end{equation}
and
\begin{equation}
    M_{\rm giant}\sin^3 i=0.18\pm0.01~M_\odot.
\end{equation} 

\subsection{Properties of the Red Giant} \label{section:giant}

\subsubsection{Modeling the spectral energy distribution (SED)}

We characterized the red giant using both fits to its overall SED and analyses of the available spectra. For the SED, we used photometry from APASS DR10 \citep{Henden2018}, Pan-STARRS DR1 \citet{Chambers2016},  2MASS \citep{Skrutskie2006} and AllWISE \citep{Wright2010AJ}. We used the \textit{Swift} \textit{UVM2} and \textit{GALEX} NUV photometry as upper limits. The compilation of the multi-band photometry used in these fits are given in Table \ref{tab:phot}. 

We initially fit single-star models to the SED of 2M0412 using DUSTY \citep{Ivezic1997,Elitzur2001} inside a MCMC wrapper \citep{Adams2015}. We assumed only foreground extinction due to $R_V=3.1$ dust \citep{Cardelli1989} and used the ATLAS9 \citep{Castelli2003} model atmospheres for the star. We assume that the source is at a distance of 3.7~kpc and used minimum luminosity uncertainties of 10\% for each band to compensate for any systematic errors and for the observed variability. We used a temperature prior based on APOGEE DR16 of $T_{\rm eff, giant}=4188 \pm75$~K and a prior of $E(B-V)=0.54\pm0.10$ on the extinction from \verb"mwdust" \citep{Bovy2016,Marshall2006,Drimmel2003,Green2019}. Ignoring the upper limits, we are fitting 11 photometric bands with three variables ($L_*$, $T_*$ and $E(B-V)$) and two priors, leading to a best fit with $\chi^2=3.8$ for 10 degrees of freedom. The single-star SED fit yields $T_{\rm eff, giant} \simeq 4207\pm68$~K, $L_{\rm giant}= 196 \pm 15~ L_\odot$, $R_{\rm giant}\simeq26.4\pm0.6~R_\odot$ and $E(B-V)= 0.587 \pm 0.046$ mag. 
%Figure \ref{sedfit} shows the SED and the best fitting model. The SED well-constrains the temperature and is consistent with the extinction estimates. 

However, given that we detect a second component in the CCF, we use \verb"VOSA" \citep{Bayo2008} and the ATLAS9 \citep{Castelli2003} model atmospheres to fit the SED of 2M0412 using a binary star model. We keep the reddening fixed at $E(B-V)=0.54$ and assume a metallicity of $\rm [M/H]=-0.5$. We obtain a good fit to the data and obtain $T_{\rm eff, giant} \simeq 4000\pm125$~K, $L_{\rm giant}=148 \pm 16~ L_\odot$, $R_{\rm giant}\simeq25.4\pm1.4~R_\odot$, $T_{\rm eff, comp} \simeq 5000\pm125$~K, $L_{\rm comp}=56 \pm 6~ L_\odot$, and $R_{\rm comp}\simeq11.1\pm0.6~R_\odot$ (Table \ref{tab:sedparams}). The two-star SED model indicates that if the companion is stellar, it is a giant located on the lower red giant branch.

\begin{figure}
	\includegraphics[width=0.5\textwidth]{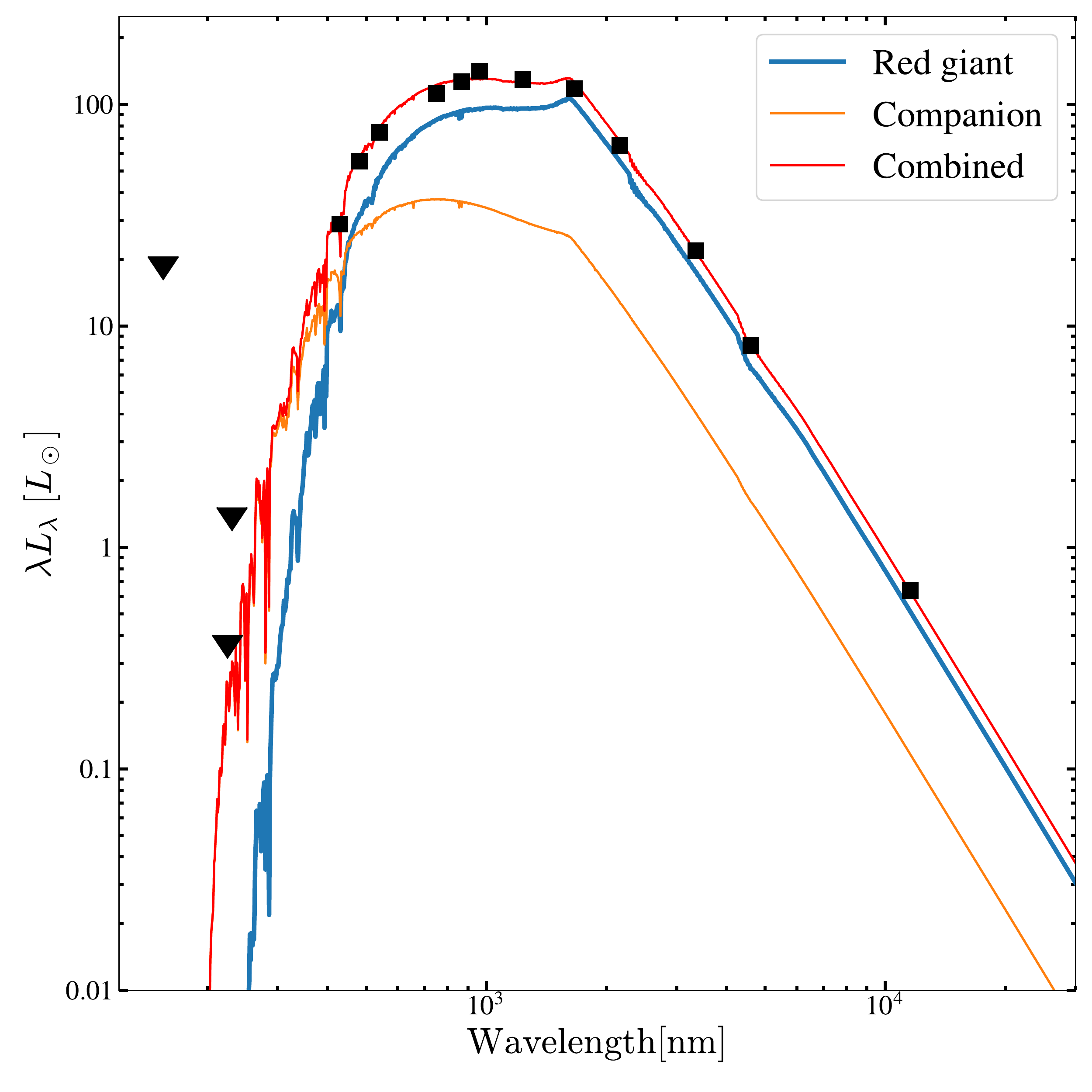}
    \caption{The best-fitting, extinction-corrected 2-star VOSA SED model for 2M0412. The multi-band photometry for 2M0412 is shown in black. The \textit{Swift} \textit{UVM2} and \textit{GALEX} \textit{FUV/NUV} upper limits are shown as arrows. The observed SED is well fit with a composite SED of a red giant and subgiant binary.}
    \label{sedfit}
\end{figure}

\begin{table*}
	\centering
	\caption{Multi-band photometry measurements used in the construction of the SED for 2M0412. Luminosities for each band are calculated assuming a nominal distance of $d\simeq3.7$~kpc. The fluxes and luminosities have not been corrected for extinction.}
	\label{tab:phot}
\begin{tabular}{rrrrrr}
		\hline
		 Filter & Magnitude & $\sigma$ & $F_\lambda\,[\rm ergs~ s^{-1}~ cm^{-2} ~ \angstrom^{-1}]$ & $\lambda L_\lambda\,[L_\odot]$ &  Reference\\
		\hline

\textit{GALEX} FUV & $>19.9$ & $5\sigma$ & $<7.1\times10^{-13}$ & $<0.296$ &  AIS; \citet{Bianchi2014} \\
\textit{GALEX} NUV & $>22.7$ & $5\sigma$ & $<2.5\times10^{-13}$  & $<0.015$   & MIS; \citet{Bianchi2014} \\
\textit{Swift} UVM2 & $>22.8$ & $3\sigma$ & $<1.0\times10^{-14}$ & $<0.004$  & This work \\
Johnson B & 16.34 & 0.05 & $1.1\times10^{-12}$ & 4.0  &  \citet{Henden2018} \\
Johnson V & 14.39 & 0.04 & $1.5\times10^{-12}$  & 16.8  &  \citet{Henden2018} \\
Pan-STARRS g & 15.15 & 0.01 & $3.5\times10^{-15}$ & 9.7   & \citet{Chambers2016} \\
Pan-STARRS r & 13.55 & 0.01 & $9.5\times10^{-15}$  & 33.4 & \citet{Chambers2016}\\
Pan-STARRS i & 13.02 & 0.01 & $1.0\times10^{-14}$  & 44.3  & \citet{Chambers2016} \\
Pan-STARRS z & 12.50 & 0.01 & $1.3\times10^{-14}$  & 59.7 &  \citet{Chambers2016}\\
Pan-STARRS y & 12.14 & 0.01 & $1.4\times10^{-14}$  & 75.7 & \citet{Chambers2016}\\
2MASS J & 10.81 & 0.02 & $1.3\times10^{-14}$ & 90.3   & \citet{Skrutskie2006} \\
2MASS H & 9.94 & 0.02 & $1.0\times10^{-14}$ & 98.1  &  \citet{Skrutskie2006} \\
2MASS $K_s$ & 9.69 & 0.02 & $4.9\times10^{-15}$  & 60.6 & \citet{Skrutskie2006} \\
WISE W1 & 9.47 & 0.02 & $1.1\times10^{-15}$  & 21.5  &  \citet{Wright2010AJ} \\
WISE W2 & 9.52 & 0.02 & $3.2\times10^{-16}$  & 8.6  &  \citet{Wright2010AJ} \\
WISE W3 & 9.37 & 0.04 & $1.0\times10^{-17}$   & 0.7  &  \citet{Wright2010AJ} \\
WISE W4 & $>8.29$ & $2\sigma$ & $<2.2\times10^{-18}$   & $<0.3$ & \citet{Wright2010AJ} \\

\hline
\end{tabular}
\end{table*}

\begin{table}
\begin{center}
\caption{Parameter estimates from the two-star SED model fit. We used model atmospheres with $\rm [M/H]=-0.5$ and assumed a reddening of $E(B-V)=0.54$.}\label{tab:sedparams}
	\begin{tabular}{r c c c }
	\hline	
	& Parameter & Companion   &   Giant \\
	\hline
	\vspace{2mm}
	& $T_{\rm{eff}}~(K)$ & $5000 \pm 125$ & $4000\pm 125$\\
	\vspace{2mm}
	& $R~(R_{\odot})$ & $11.1\pm0.6$ & $25.4\pm1.4$ \\	
	\vspace{2mm}
	& $L~(L_{\odot})$ & $56\pm6$  & $148\pm16$ \\	
	\hline
	\vspace{2mm}	
\end{tabular}
\end{center}
\end{table}

\subsubsection{Modeling the HIRES/PEPSI spectra}

The APOGEE, HIRES and PEPSI spectra indicate that the giant is rapidly rotating. The APOGEE DR16 spectra were used to measure $v_{\rm rot} \, \textrm{sin} \, i=12.61\pm0.92 \, \rm km \,s^{-1}$ following the methods described in \citet{Tayar2015,Dixon2020,MazzolaDaher2021}. The HIRES CPS pipeline \citep{Petigura2015} reports $v_{\rm rot} \, \textrm{sin} \, i=13.3\pm1.0 \, \rm km \,s^{-1}$ and we found $v_{\rm rot} \, \textrm{sin} \, i=12.8 \pm 1.2 \, \rm km \,s^{-1}$ from the PEPSI spectrum using \verb"iSpec" \citep{Blanco-Cuaresma2014,Blanco-Cuaresma2019}. 

Assuming that the rotation of the giant is tidally synchronized with the orbit, we can derive the rotational velocity  \begin{equation}
      v_{\rm rot} = {2\pi R_{\rm giant} \over P_{\rm rot}}=15.6\pm0.4\, \rm km \,s^{-1}~\bigg(\frac{R_{\rm giant}}{25~R_\odot}\bigg)
\end{equation} for the giant. This is a reasonable assumption because binaries with evolved components ($\log(g)<2.5$) and orbital periods shorter than ${\sim}100$~days are expected to be tidally locked (e.g., \citealt{Verbunt1995,Price-Whelan2018}). However, the light curve residuals after removing the ellipsoidal variability show no periodic signal that might confirm or reject this hypothesis. The APOGEE DR16, HIRES and PEPSI measurements combined with the period and stellar radius yield estimates of $\sin\, i=0.80\pm0.05$, $\sin\, i=0.85 \pm 0.05$ and $\sin\, i=0.82 \pm 0.06$, respectively, and orbital inclinations of $i=53.1\pm3.9^{\circ}$, $i=58.2\pm4.1^{\circ}$ and $i=55.1\pm4.7^{\circ}$, respectively. To investigate the rotation of the second component, we subtract a synthetic spectrum of the giant from the HIRES spectra and shift the residuals to the companion's rest-frame. We then calculate the median spectrum from these residuals and measured $v_{\rm rot} \, \textrm{sin} \, i=17.6 \pm 0.6 \, \rm km \,s^{-1}$ for the companion.

\begin{figure*}
	\includegraphics[width=0.75\textwidth]{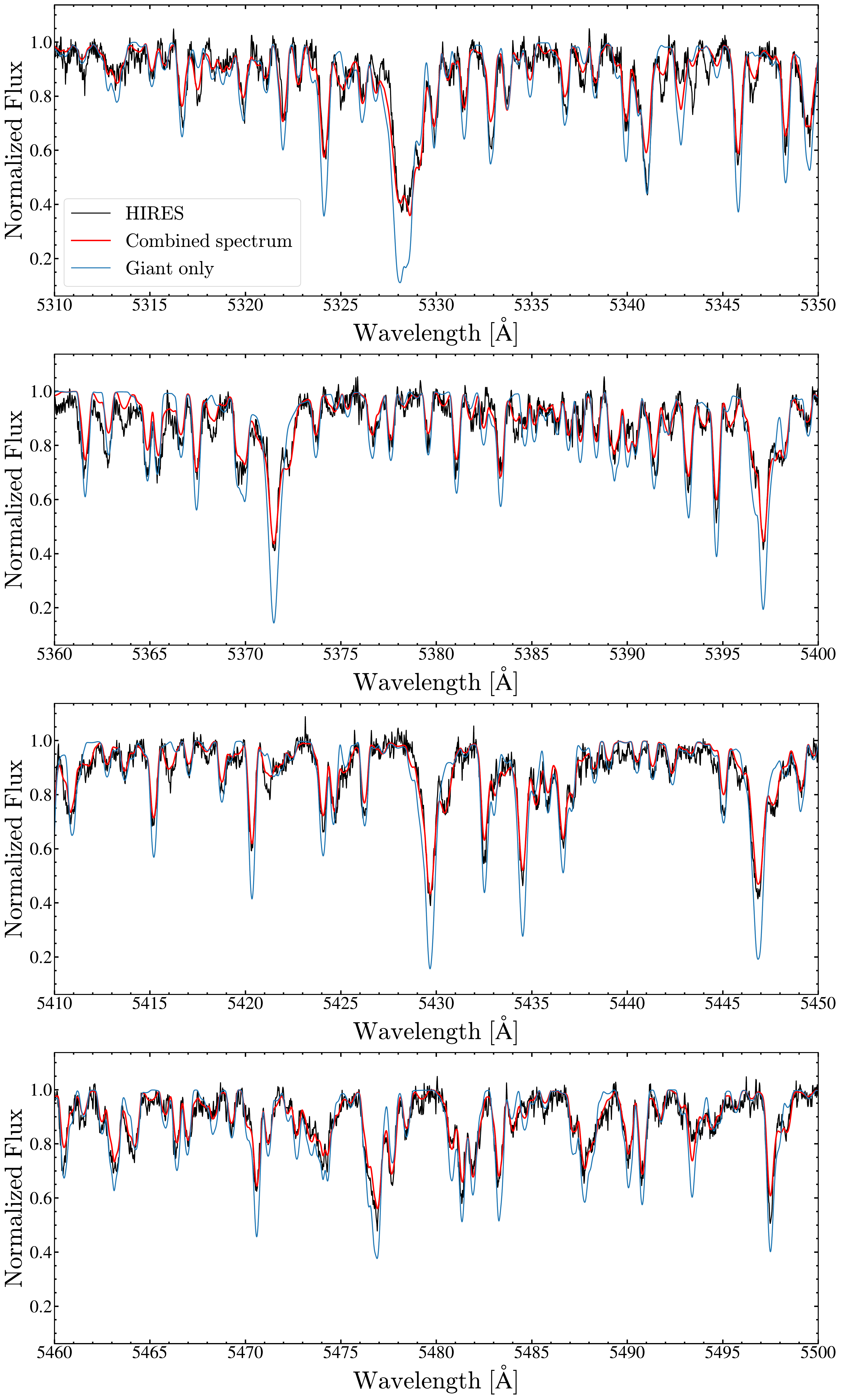}
    \caption{The Keck/HIRES spectrum (black) and the best-fit composite spectrum (red) for a red giant + lower giant binary with $F_{\rm comp}/F_{\rm tot}=0.35$. The spectrum of a single giant star is shown in blue. The composite spectrum fits the HIRES spectrum better than a single giant.}
    \label{specbin}
\end{figure*}

We use the spectral synthesis code \verb"iSpec" to synthesize spectra for a giant with $T_{\rm eff, giant} \simeq 4000$~K, $\log(g)_{\rm giant}\simeq1.2$, and a lower giant with $T_{\rm eff, comp} \simeq 5000$~K and $\log(g)_{\rm comp}\simeq2.7$ for a metallicity of $\rm [Fe/H]=-0.5$ using MARCS model atmospheres \citep{Gustafsson2008}. Our two-star SED model indicates that the subgiant companion contributes ${\sim}35\%$ of the total flux in the $V$-band between 530-550~nm, so we kept the flux ratio fixed at $F_{\rm comp}/F_{\rm tot}=0.35$ for this wavelength range. We assumed an $\alpha$-enhancement of $\rm [\alpha/Fe]=0.2$. The spectra were broadened based on our $v_{\rm rot} \, \textrm{sin} \, i$ measurements for the giant and the companion. We shifted the spectrum of the companion based on the derived velocities and then combined the two spectra using the flux ratio indicated by the binary SED model. Figure \ref{specbin} shows the combined synthetic spectrum and the observed HIRES spectrum at phase $\phi\simeq0.68$. The combined spectrum (Figure \ref{specbin}) is a very good fit to the data. We also note that the spectrum of a giant without additional veiling and absorption lines from a companion cannot fit the data as well as the combined model.

\begin{figure}
	\includegraphics[width=0.5\textwidth]{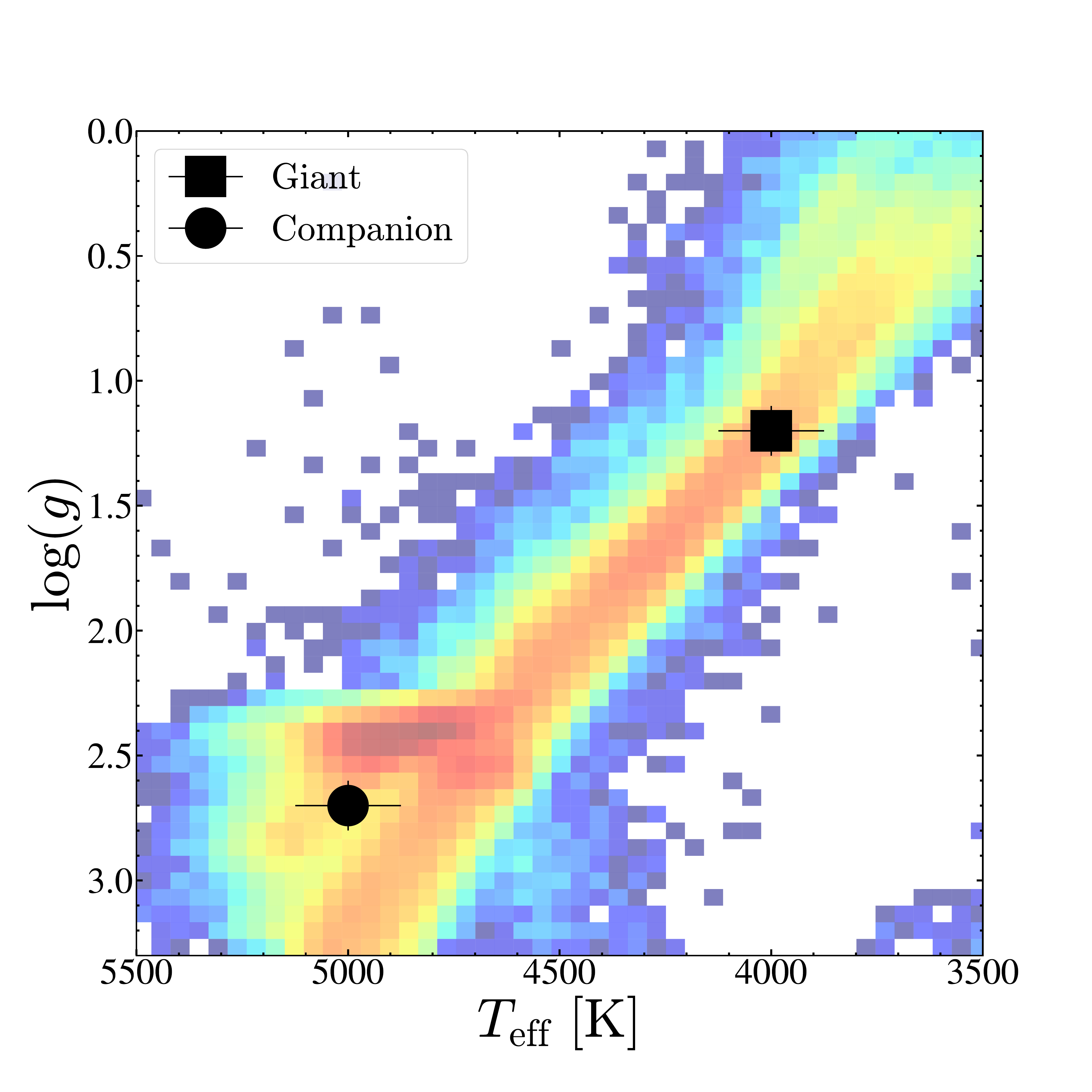}
    \caption{$\log(g)$ vs. $T_{\rm eff}$ for the sample of APOGEE DR16 giants with $-1<\rm[M/H]<0$. The positions of the giant and the companion are shown in black.}
    \label{loggteff}
\end{figure}

Figure \ref{loggteff} shows the position of the giant and the companion in $\log(g)$ vs. $T_{\rm eff}$. Both the giant and the companion have $T_{\rm eff}$ and $\log(g)$ that agree with the distribution of APOGEE DR16 sources. In DR16, 2M0412 has $T_{\rm eff}=4187\pm75$, $\log(g)=1.80\pm0.06$ and $\rm [Fe/H]=-0.73\pm0.01$. The measured value of $\log(g)$ in DR16 is somewhat larger than our estimate. However, the APOGEE ASPCAP pipeline routinely overestimates $\log(g)$ for rapidly rotating giants because the ASPCAP template library of synthetic spectra does not include rotation for giants. For giants rotating at $v_{\rm rot} \, \textrm{sin} \, i\sim 8-14 \, \rm km \,s^{-1}$, the offset in $\log(g)$ can be as large as 0.1-0.5 dex (see, e.g., \citealt{Thompson2019}). This bias can also affect other spectroscopically determined parameters for these rapidly rotating stars. Given that the companion contributes ${\sim}20\%$ of the light in the APOGEE spectra, the single star ASPCAP fits will also yield biased results. \citet{ElBadry2022} presents binary fits to the APOGEE spectra that are consistent with our results from fitting the spectra with a two-star model.

\subsection{Modelling the ellipsoidal variability with PHOEBE}\label{section:phoebe}

We interpret the light curves as ellipsoidal variability and fit models to derive the orbital properties and dynamical masses of the red giant and its companion. 

We tried MCMC fits of the ZTF $g$-band, $r$-band and TESS light curves using \verb"PHOEBE" \citep{Phoebe2016,Horvat2018,Conroy2020, Foreman-Mackey2013} with priors on the radius ($R_{\rm giant}=25.4 \pm 1.5 R_\odot$). We do not consider the effects of irradiation and reflection in our models. We adopt a logarithmic limb-darkening law and use coefficients from \citet{Claret2011} and \citet{Claret2017}. We fixed the orbital parameters to those obtained from \verb"yorbit" (Table \ref{tab:orbit}) and kept the mass ratio fixed at $q=0.201$. Initial models suggested that the giant was close to filling its Roche lobe and spectroscopic evidence for an accretion disk provides additional support for the argument that the giant is filling its Roche Lobe (see $\S$\ref{section:balmer}), so we constrained the giant to be semi-detached. We keep the radius of the companion and the temperatures fixed at $R_{\rm comp}\simeq 11.1 R_\odot$, $T_{\rm eff, giant}=4000$~K and $T_{\rm eff, comp}=5000$~K. We fit for the orbital inclination ($i$), and the radius of the giant ($R_{\rm giant}$). We marginalize over the passband luminosities and nuisance parameters that account for underestimated errors in the light curves.

\begin{figure*}
	
	\includegraphics[width=0.7\textwidth]{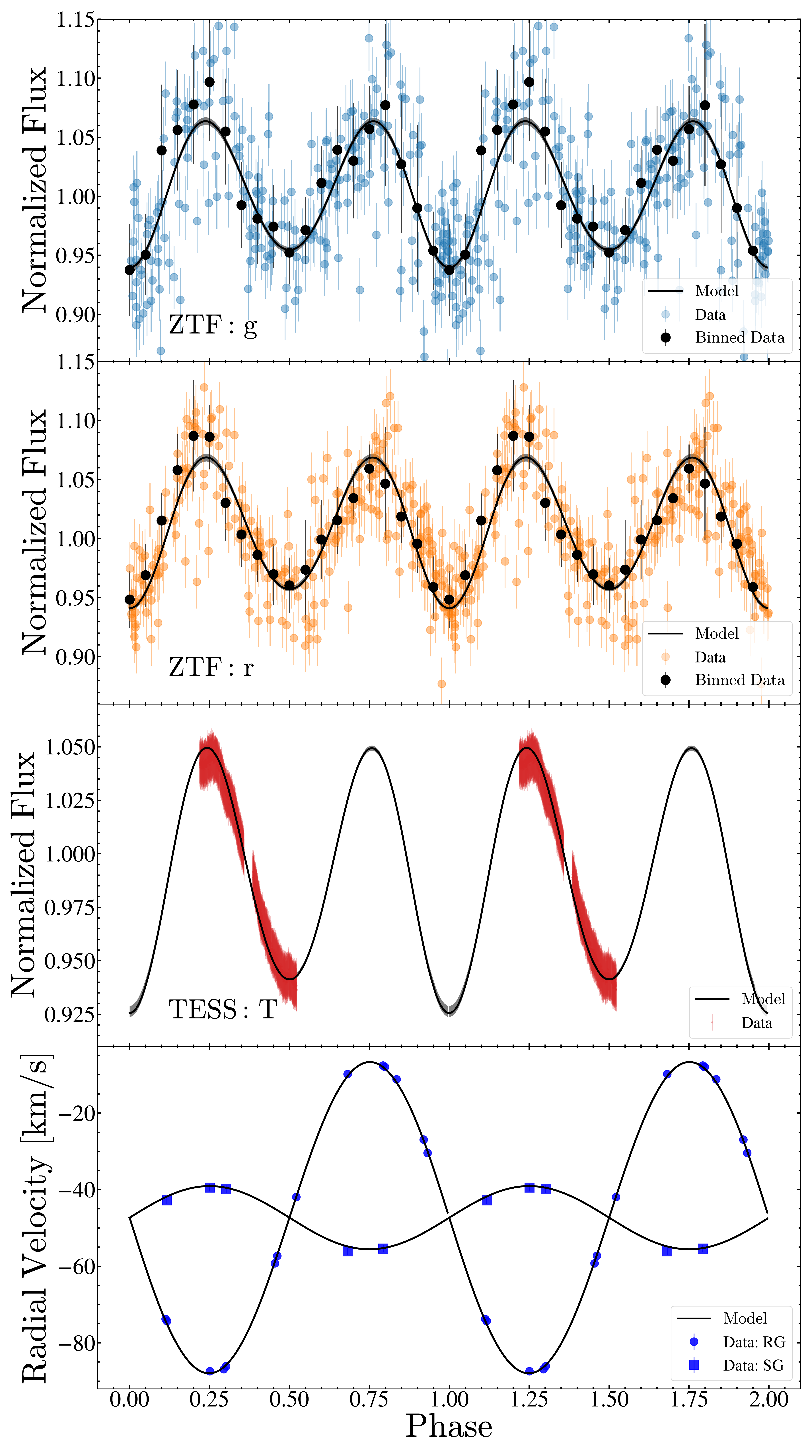}
		\vspace{-0.2cm}	
    \caption{The normalized ZTF $g$, ZTF $r$ and TESS $T$-band light curves for 2M0412 as a function of orbital phase (defined with the epoch of maximal RV at $\phi=0.75$). The binned light curves for the ZTF $g$ and ZTF $r-$ band light curves are shown as black points. The scatter in the bins are shown as errorbars. The light curves and RV curves from the best fitting \textsc{PHOEBE} model are also shown.}
    \label{lcs}
\end{figure*}

Table \ref{tab:params} presents the results of our best fitting \verb"PHOEBE" model. The reported errors are purely statistical and are likely an underestimate of the true uncertainty in the model. This is because we do not consider systematic effects in the derivation of the binary solution. The mass of the giant is $M_{\rm giant}=0.38^{+0.01}_{-0.01}~M_\odot$, the radius of the giant is $R_{\rm giant}=26.1\pm0.1~R_\odot$ and the orbital inclination is $i_{\rm orb}=49.44{^{+0.31}_{-0.36}}^\circ$. The radius of the giant derived from the \verb"PHOEBE" model is consistent with that obtained from the two-star SED model in $\S$\ref{section:giant} (Table \ref{tab:sedparams}). The inferred mass of the red giant ($M_{\rm giant}\simeq0.38~M_\odot$) is inconsistent with single-star evolution and instead implies that it has been heavily stripped by binary interaction. In this model, the companion mass is $M_{\rm comp}=1.91^{+0.03}_{-0.03}~M_\odot$. This model is a good fit to the ZTF $g$-band, $r$-band, the \textit{TESS} $T$-band light curve, and the radial velocities (Figure \ref{lcs}). 

\begin{table}
\begin{center}
\caption{PHOEBE parameter estimates for the red giant and companion. The reported errors are purely statistical and do not include systematic effects  }\label{tab:params}
	\begin{tabular}{r c c c }
	\hline	
	& Parameter &  Companion  &    Giant\\
	\hline
	\vspace{2mm}
	& $P_{\rm orb}$  (d) & \multicolumn{2}{c}{81.1643 (fixed) } \\
	\vspace{2mm}
	& $\omega~(^\circ)$ & \multicolumn{2}{c}{$0$ (fixed)}\\% & $-$ & $-$\\
	\vspace{2mm}
	& $e$ & \multicolumn{2}{c}{$0$ (fixed)}\\%& $-$ & $-$\\
	\vspace{2mm}
	& ${\gamma}~(\rm km ~s^{-1})$ &  \multicolumn{2}{c}{-47.031 (fixed)}\\% & $-$ & $-$ \\
	\vspace{2mm}
	& $a ~(R_{\odot})$ & $17.4 ^{+0.1}_{-0.1}$ & $86.7^{+0.5}_{-0.4}$ \\
	\vspace{2mm}	
	& $i~(^\circ)$  & \multicolumn{2}{c}{$49.44^{+0.31}_{-0.36} $ }\\% &  $-$ & $-$ \\
	\vspace{2mm}	
	& $T_{\rm{eff}}~(K)$ & $5000$ (fixed) & $4000$ (fixed)\\
	\vspace{2mm}
	& $R~(R_{\odot})$ & $11.1$ (fixed)  & $26.1\pm0.1$\\	
	\vspace{2mm}
	& $q$ & \multicolumn{2}{c}{$0.201$ (fixed)} \\% $-$ & $-$\\
	\vspace{2mm}	
	& $M~(M_{\odot})$ & $1.91\pm0.03$ & $0.38\pm0.01$\\
	\vspace{2mm}	
	& $\log(g)$ & $2.64\pm0.01$ & $1.19\pm0.01$\\	
	\hline
	\vspace{2mm}	
\end{tabular}
\end{center}
\end{table}

%\clearpage
%\clearpage

\subsection{Balmer $\rm H\alpha$ emission} \label{section:balmer}

\begin{figure}
    
    \includegraphics[width=0.5\textwidth]{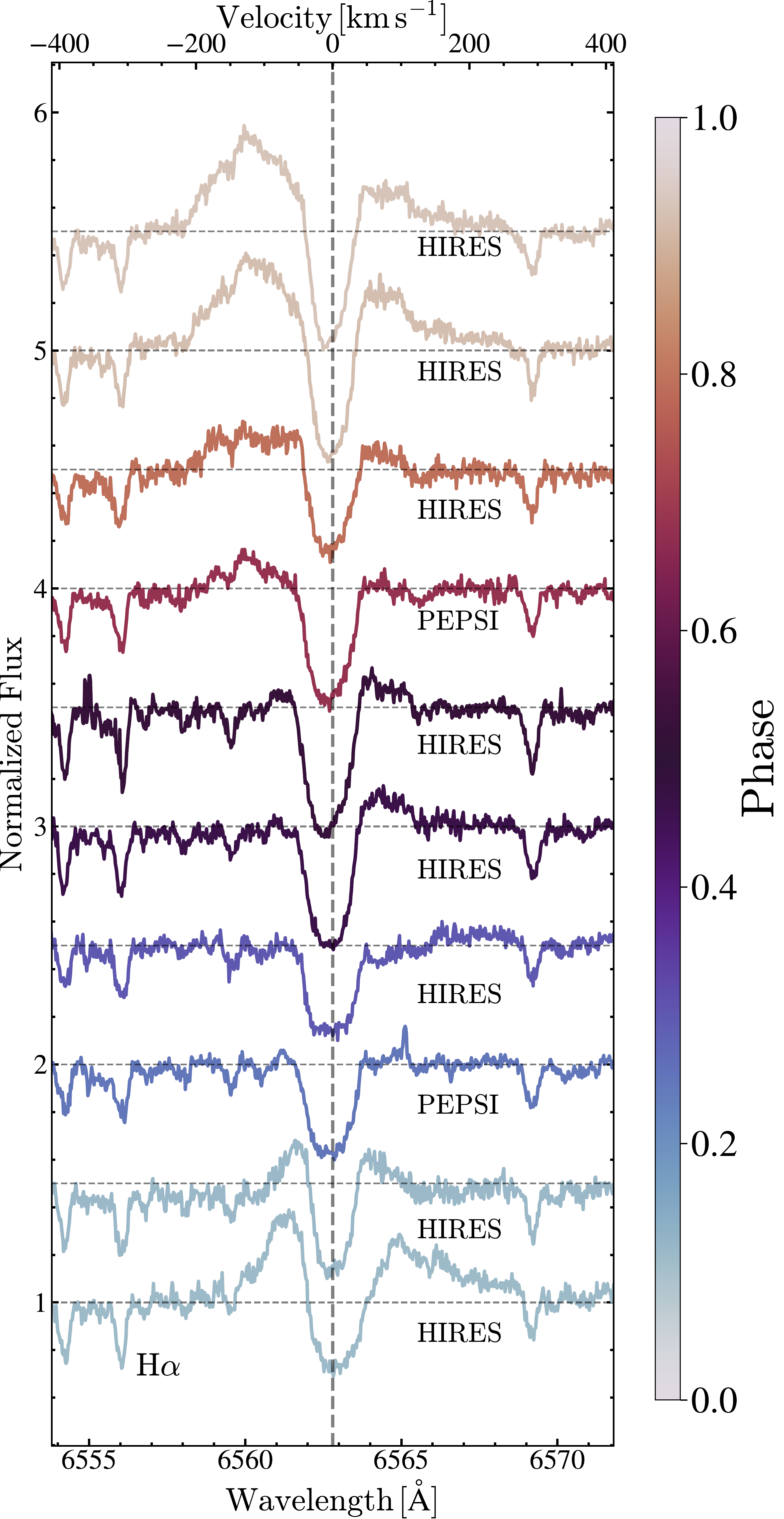}
    \caption{$\rm H\alpha$ line profiles in the rest frame of the giant in the HIRES and PEPSI spectra. The line profiles are sorted and colored as a function of orbital phase.}
    \label{spechalpha}
\end{figure}

As shown in Figure \ref{spechalpha}, there is a broad $\rm H\alpha$ emission line whose structure varies with the orbital phase. V723 Mon also has this property \citep{Jayasinghe2021}, but here it is directly visible without any need to subtract the stellar spectrum.

We identify both blue-shifted and red-shifted emission components in most of our spectra. We define these components by the wavelength ranges of $6560-6562~\angstrom$ and $6564-6566~\angstrom$ respectively. The most significant $\rm H\alpha$ emission in our spectra is seen near phase $\phi{\sim}0.9$ as a blue-shifted component. The emission component has a velocity close to the rest frame of the giant near phase $\phi{\sim}0.25$ and fills in the giant's $\rm H\alpha$ absorption line. Near phase ${\sim}0.9$, the typical Gaussian FWHM of the blue-shifted (red-shifted) $\rm H\alpha$ emission component is ${\sim}255\, \rm km\,s^{-1}$ (${\sim}265\, \rm km\,s^{-1}$). The wings of the emission extends beyond ${\sim}200\, \rm km\,s^{-1}$ in some of our spectra. 

The peak separations at $\phi{\sim}0.1$ and $\phi{\sim}0.9$ are ${\sim}160\, \rm km\,s^{-1}$ and ${\sim}196\, \rm km\,s^{-1}$ respectively. However, the behavior of the double-peaked profile appears to vary between orbits. For example, we obtained Keck/HIRES spectra near $\phi{\sim}0.1$ in two consecutive orbits, where the first spectrum has a peak separation of ${\sim}106\, \rm km\,s^{-1}$ and the second spectrum has a much larger peak separation of ${\sim}160\, \rm km\,s^{-1}$. For the epochs where we can see a double-peaked emission profile, the median peak separation is ${\sim}168\, \rm km\,s^{-1}$ with a scatter of $\sigma{\sim}31\, \rm km\,s^{-1}$. 

\begin{figure}
	\includegraphics[width=0.5\textwidth]{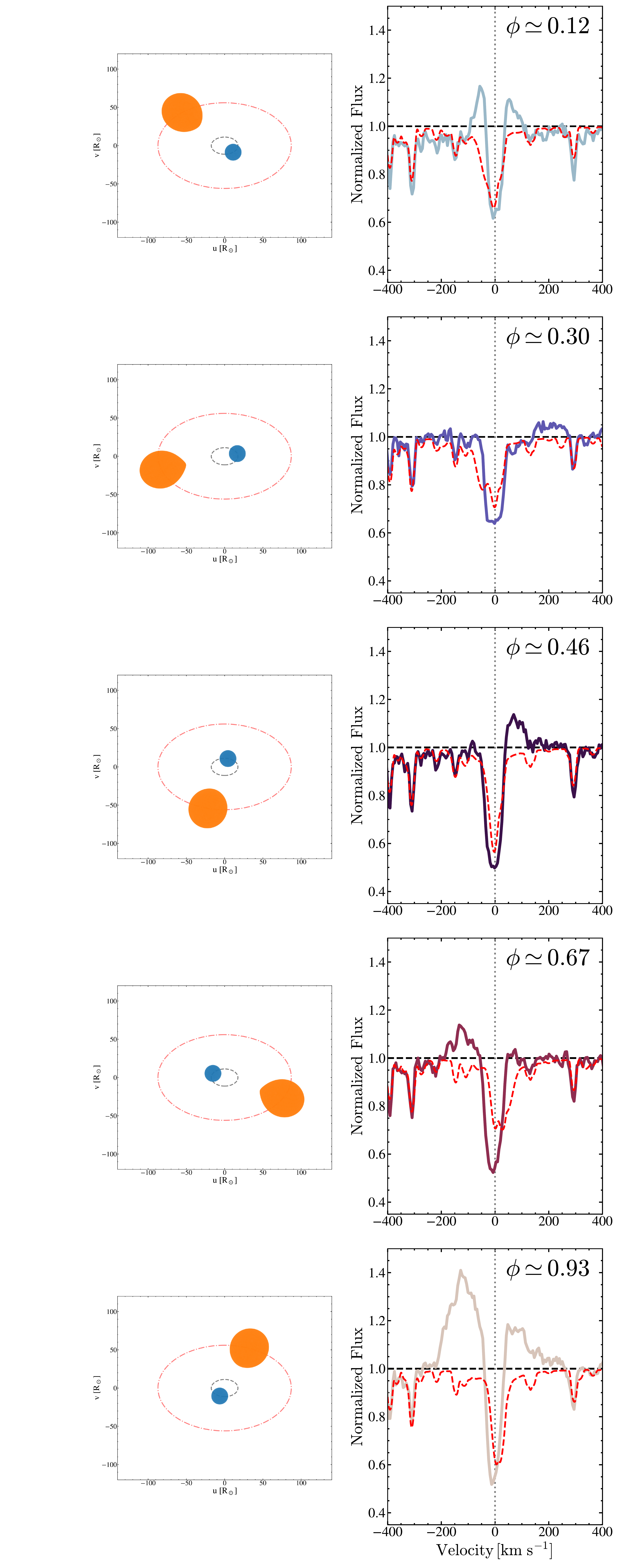}
	
    \caption{Behavior of the $\rm H\alpha$ line profiles with orbital phase. The right panels show the $\rm H\alpha$ line profiles at the orbital configurations illustrated in the left panels. The red dashed line shows the synthetic composite spectrum for the binary at the corresponding orbital phase.}
    \label{orbhalpha}
\end{figure}

\begin{figure*}
	\includegraphics[width=\textwidth]{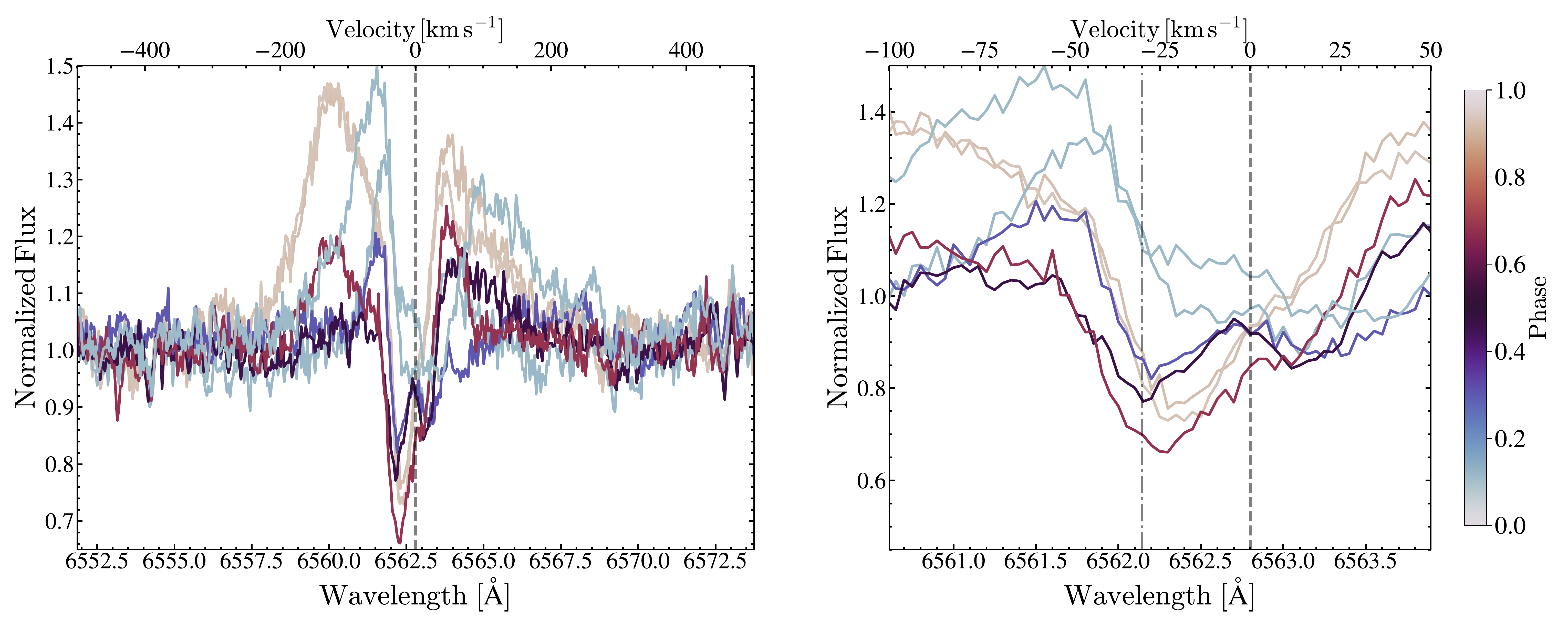}
	
    \caption{Variations of the $\rm H\alpha$ line profiles with orbital phase in the giant's rest frame after subtracting a template composite spectrum for the binary. The left panel shows the full wavelength range and the right panel zooms in on the line core. Note the $\rm H\alpha$ absorption feature at ${\simeq}-30\, \rm km\,s^{-1}$ in the giant's rest-frame.}
    \label{halphasub}
\end{figure*}

\begin{figure*}
    
    \includegraphics[width=1\textwidth]{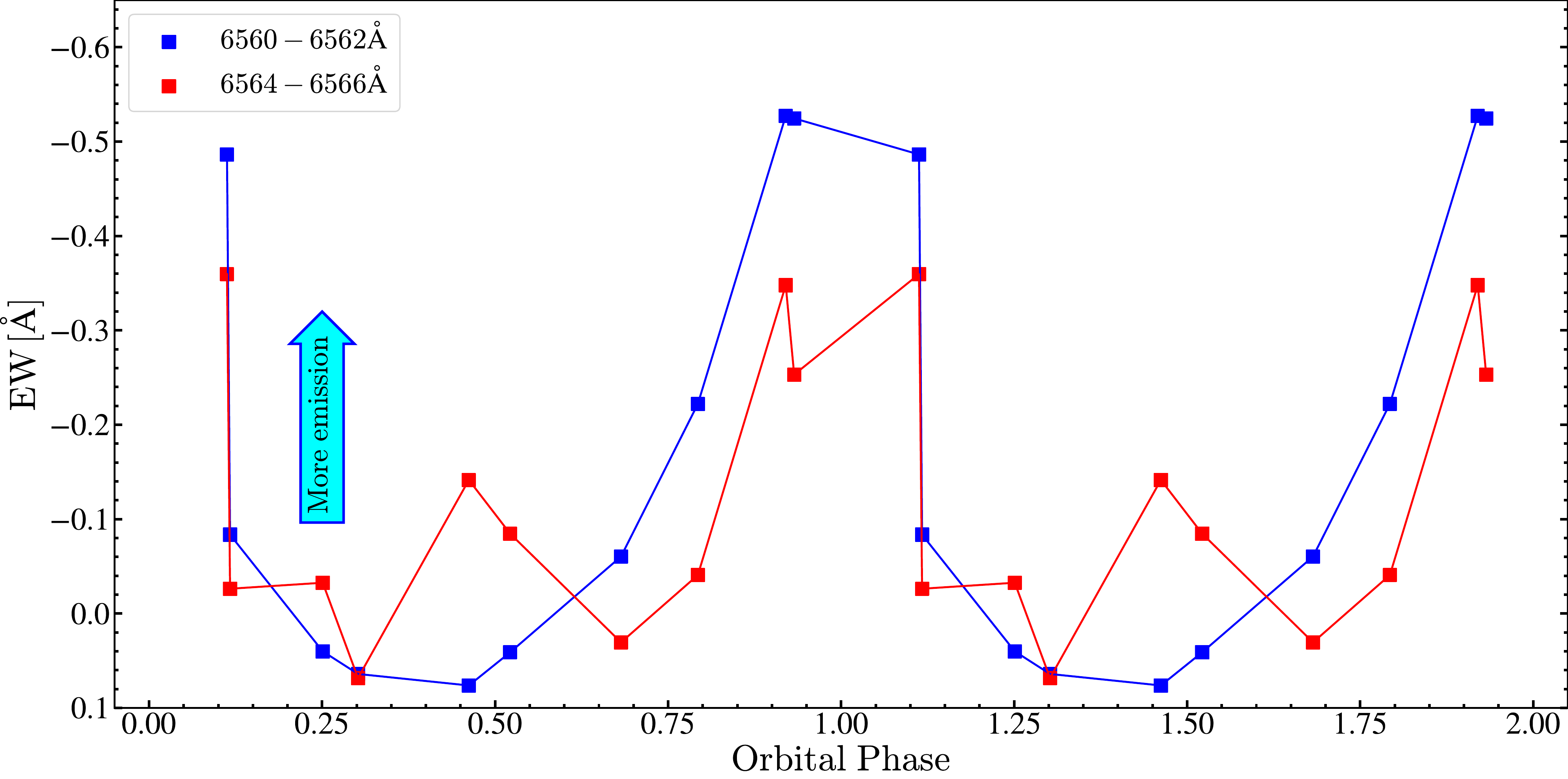}
    \caption{Equivalent width of the blue-shifted (blue points, $6560-6562~\angstrom$) and red-shifted (red points, $6564-6566~\angstrom$) $\rm H\alpha$ emission components as a function of orbital phase.}
    \label{ewphase}
\end{figure*}

Figure \ref{orbhalpha} shows the $\rm H\alpha$ line at various orbital phases along with the  \verb"PHOEBE" models of the orbital configurations of the binary. After conjunction at $\phi\simeq0.5$, the emission is blue-shifted and reaches a maximum at phase $\phi{\sim}0/1$ (Figure \ref{ewphase}). Since the $\rm H\alpha$ emission varies significantly in the rest-frame of the giant, this emission must emerge from beyond the giant's photosphere. Given its large FWHM, we can rule out a chromospheric origin for the $\rm H\alpha$ emission. We also do not observe the $\ion {Ca}{ii}$ H and K emission lines that are typically associated with chromospheric activity,

Figure \ref{halphasub} shows the $\rm H\alpha$ line profiles after subtracting the template spectrum of the red giant. We see a blue-shifted $\rm H\alpha$ absorption feature in these line profiles at a roughly constant ${\sim}30\, \rm km\,s^{-1}$ blue shift relative to the giant. The depth of the absorption core is strongest near $\phi{\sim}0.5$. The velocity of this absorption feature is close to the velocity of the giant's photosphere ($v_{\rm rot, giant}\simeq 16\, \rm km\,s^{-1}$), consistent with mass outflow from the giant at $L_1$. At $\phi=0$, the inner Lagrangian point ($L_1$) is directed toward the observer and at $\phi=0.5$, $L_1$ points away from the observer \citep{Beech1985}. It is possible that the absorption feature is associated with the companion or density enhancements caused by matter streaming through the inner Lagrangian point. 

Figure \ref{ewphase} shows how the equivalent width varies with orbital phase for both the blue-shifted and red-shifted $\rm H\alpha$ components. The largest equivalent width of the blue-shifted (red-shifted) $\rm H\alpha$ component is $\rm EW(H\alpha){\sim}-0.5$ ($\rm EW(H\alpha){\sim}-0.3$) at $\phi{\sim}0.9$. These equivalent widths are more than an order of magnitude smaller than those observed for X-ray binaries (see, for e.g., \citealt{Fender2009}) and correspond to a total line luminosity of ${\sim}0.003~L_\odot$. For both the blue and red components, the equivalent widths are the most negative (emission is strongest) near $\phi{\sim}0$ when the companion is on the near side of the giant. The red-shifted line also has a weaker peak at $\phi{\sim}0.5$ when the companion is on the far side.

If we interpret the double peaked structure as an accretion disk, the projected radial velocity of the outer rim of an accretion disk is approximately equivalent to half the velocity separation of the two peaks $V_h$ (see, for e.g., \citealt{Smak1981,Strader2015}). In this case, the $V_h$ of $84\, \rm km\,s^{-1}$ is significantly larger than the radial velocity semi-amplitude of the red giant (${\sim}42\, \rm km\,s^{-1}$). It is also slightly larger than the circular velocity at the surface of the giant (${\sim}78\, \rm km\,s^{-1}$ for $M=1~M_\odot$ and $R=31~R_\odot$). Indeed, $V_h$ is expected to be larger than the radial velocity semi-amplitude of the giant for classical Keplerian disks (see, for e.g., \citealt{Paczynski1977,Orosz1994}). 
%As an orbit around the secondary, this velocity corresponds to ${\simeq}27~M_{\rm comp}\sin^2i ~R_\odot$ or ${\simeq}16~M_{\rm comp} ~R_\odot$ if we use a disk inclination of $50^\circ$.

In summary, we observe a phase dependent, broad ($\gtrsim200\, \rm km\,s^{-1}$)  $\rm H\alpha$ emission line in our spectra of 2M0412. We show that the morphology and orbital evolution of the $\rm H\alpha$ emission is likely due to an accretion disk. This is consistent with the red giant filling its Roche lobe. However, more spectra are needed to study the behavior of the $\rm H\alpha$ emission in greater detail.

\subsection{X-ray upper limit} \label{section:xrays}

No X-rays were detected in the \textit{Swift} XRT observations. We derived a 3 sigma upperlimit using the reprocessed observations of sw00014417001 (${\sim}2.2$~ks), a 50 arcsec source region centered on the source and a 150 arcsec source free background region located at $(\alpha_{J2000}, \delta_{J2000})$=(04:12:38.10,67:43:00.86). We derive an aperture corrected 3$\sigma$ upper limit on the 0.3-2.0 keV count rate of $8.1\times10^{-4}$ counts/s. Using the Galactic column density along the line of sight ($N_{H}\sim1.34\times10^{22}$ cm$^{-2}$, \citealt{HI4PICollaboration2016}), and an absorbed power-law with a photon index of 2, we derive an absorbed (unabsorbed) flux of $1.7\times10^{-14}$ erg cm$^{-2}$ s$^{-1}$ ($5.0\times10^{-14}$ erg cm$^{-2}$ s$^{-1}$), which corresponds to an absorbed (unabsorbed) X-ray luminosity of $2.9\times10^{31}$ erg s$^{-1}$ ($8.1\times10^{31}$ erg s$^{-1}$). The \textit{Swift} XRT limit on the X-ray luminosity is comparable to the $L_{\rm X}\sim10^{30}-10^{31}~ \rm ergs~s^{-1}$ observed for quiescent X-ray binaries \citep{Dincer2018} and the $L_{\rm X}\sim10^{29}-10^{30}~ \rm ergs~s^{-1}$ observed for chromospherically active RS CVn systems \citep{Demircan1987}. 

%If the X-ray luminosity originates from an accretion disk around a ${\sim}3.0~M_\odot$ black hole, the luminosity is ${\lesssim}1.1\times10^{-7}~L_{\rm edd}$. 

\section{Comparison to binary evolution models} \label{section:bpass}

We used \verb"hoki" \citep{Stevance2020,Stevance2021} to search for Binary Population And Spectral Synthesis (BPASS; \citealt{Eldridge2017,Stanway2018}) binary models consistent with the observed properties of 2M0412. We used the BPASS v2.2.1 models at $\rm Z=0.006$, which is the closest match to the metallicity derived in $\S$\ref{section:giant}. 

We restrict our search to models with
\begin{enumerate}
  \item Period of the binary, $P_{\rm orb}=81\pm8$~days,
  \item $V=-0.6\pm1$~mag,
  \item $J=-2.8\pm1$~mag,
  \item $K=-3.5\pm1$~mag,
  \item Total luminosity of the binary, $ \log(L_{\rm binary})=2.31\pm0.05$, 
  \item Radius of the giant, $ \log(R_{\rm giant})=1.40\pm0.05$, 
  \item Effective temperature of the giant, $\log(T_{\rm eff})=3.60\pm0.05$, 
  \item Mass of the giant, $M_{\rm giant}=[0.25,2]~M_\odot$, and
  \item Binary mass ratio, $0.15<q<0.35$.
\end{enumerate}

We searched for models where either (1) the companion is a star with an effective temperature lower than $T_{\rm eff, comp}<7000$~K, or (2) the companion is a compact object. We found 16 giant + star and 21 giant + compact object models that satisfied these criteria. The existence of numerous giant + compact object binary models in BPASS suggests that the parameter space in which 2M0412 currently exists is viable for the discovery of compact objects, though the detection of a luminous secondary leads us to reject a compact object model for 2M0412. In all of the BPASS models, the giant is heavily stripped with masses in the range of ${\sim}0.35-0.46~M_\odot$ ($M_{\rm init}\simeq1.2-1.8~M_\odot$) and ${\sim}0.35-0.79~M_\odot$ ($M_{\rm init}\simeq0.9-2.5~M_\odot$) for the giant + star and the giant + compact object models, respectively. Table \ref{tab:bpass} gives three examples of the giant + star binary models closest in period and magnitude ($P_{\rm orb}=81\pm2$~days, $\Delta(M)=0.5$~mag) to the observations.

The giant + star binary that best matches our data has $M_{\rm giant}\simeq0.35~M_\odot$, $M_{\rm comp}\simeq1.55~M_\odot$, $q\simeq0.227$, $R_{\rm giant}\simeq26.7~R_\odot$, and $T_{\rm eff, giant}\simeq4279~\rm K$. The current configuration of the giant + star binary is short-lived, with the orbit expanding to a period of ${\sim}85.7$~days in ${\sim}2.4$~Myr. In this model, the binary is undergoing RLOF for a total of ${\sim}24$~Myr. The two other giant + star binary models in Table \ref{tab:bpass} have similar properties. In these models, the giant loses ${\sim}70-80\%$ of its initial mass. For a conservative mass-transfer binary, the accretor is spun up and will therefore be rapidly rotating. The companion to the giant is a main sequence star that is ${\sim}3-4$~mag fainter than the overall binary in the $V$-band. This is inconsistent with our model for this binary, where the companion is a lower giant that contributes significantly in the $V$-band (${\sim}35\%$).

\begin{table*}
	\centering
	\caption{Properties of exemplary BPASS giant+star binary models that match observations for 2M0412.}

\begin{threeparttable}	
\begin{tabular}{rrrrrrrrrrrrrrr}
		\hline
		\vspace{1mm}	
		 Model & $M_{\rm giant}$ & $M_{\rm He~core}$ & $M_{\rm comp}$ & $q$ & $T_{\rm eff, giant}$ & $T_{\rm eff, comp}$ & $R_{\rm giant}$ & $R_{\rm comp}$  & $L_{\rm giant}$ & $P_{\rm orb}$ & Age\\
		  & $(M_{\odot})$ & $(M_{\odot})$ & $(M_{\odot})$ & & $(\rm K)$ & $(\rm K)$ & $(R_{\odot})$ & $(R_{\odot})$ & $(L_{\odot})$ &  $(\rm d)$ &  Gyr\\		 
		\hline
        1 & 0.41 & 0.34 & 2.30 & 0.176 & 4078 & 6353 & 27.8 & 2.1 & 191  & 79.3 & 2.2 \\
        2 & 0.37 & 0.33 & 2.14 & 0.173 & 4101 & 6037 & 27.5 & 2.1 & 193  & 82.9 & 2.7\\
        3 & $^{\rm a}~$0.35 & 0.33 & 1.55 & 0.227 & 4279 & 6489 & 26.7 & 1.4 & 215 & 82.0 & 1.5\\        
		\hline
\end{tabular}
    \begin{tablenotes}
      \small
      \item $^{\rm a}$ Best matching giant + star model.
   
    \end{tablenotes}

\end{threeparttable}
\label{tab:bpass}
\end{table*}

We were puzzled by (in particular) the low temperatures of the stellar companions in the BPASS models. The cause is that the present BPASS models only evolve one star at a time for reasons of speed \citep{Eldridge2016}. This can lead to misestimates of the properties of the accreting star (Eldridge, private comm.). To examine this, we evolved the BPASS giant + star binaries (Table \ref{tab:bpass}) using Modules for Experiments in Stellar Astrophysics \citep[MESA][]{Paxton2011, Paxton2013, Paxton2015, Paxton2018, Paxton2019}. MESA computes Roche lobe radii in binary systems using the fit of \citet{Eggleton1983}, and the mass transfer rates in Roche lobe overflowing binary systems are determined following the prescription of \citet{Kolb1990}. We adopt the explicit mass transfer calculation scheme in MESA. 

We use the initial component masses and orbital period ($P_{\rm orb,init}=8$~days) from the BPASS models and examine their properties when the binary has a period of $P_{\rm orb}{\sim}81$~days. As in the BPASS models, the initially more massive component becomes a red giant with $M_{\rm giant}\simeq0.32-0.34~M_\odot$ and its companion is a main sequence star with $M_{\rm comp}\simeq1.6-2.1~M_\odot$. In the MESA models, however, the companion to the giant is a hot, luminous main sequence star with $T_{\rm eff, comp}\simeq10700-11200$~K, $R_{\rm comp}\simeq1.4-1.6~R_\odot$, $\log(g)\simeq4.4$ and $L_{\rm comp}\simeq24-28~L_\odot$. Such a hot companion is grossly incompatible with the NUV limits from \textit{Swift} and \textit{GALEX} ($\S$\ref{section:giant}, Figure \ref{sedfit}). This is also significantly hotter than the temperature derived for the companion from fitting the SED and the spectra ($T_{\rm{eff, comp}}=5000\pm125$~K). The higher temperatures in the {\tt MESA} models would make the companion easily visible. For the companion to be a lower giant with $T_{\rm{eff, comp}}\simeq 5000$~K, the initial configuration of the binary has to be more finely tuned than the BPASS models that we have explored in this work, such that the secondary is already leaving the main-sequence when mass transfer begins. A detailed exploration of such MESA models is presented in \citet{ElBadry2022}.

%For all three models, the giant is still in the process of transferring mass to its companion when it reaches this period. %The giant's properties of $T_{\rm eff, giant}\simeq4094-4103$~K, $R_{\rm giant}\simeq28.7-30.2~R_\odot$, $\log(g)\simeq1.3$ and $L_{\rm giant}\simeq209-232~L_\odot$ are virtually identical to those of the observed star. For all three models, the giant is still in the process of transferring mass to its companion when it reaches this period. Once the mass transfer is complete, the binary is at an orbital period of $P_{\rm orb}\simeq220-281$~days.

\section{Discussion}

\label{section:darkcomp}

In this section, we systematically discuss the nature of the companion given the observed properties of the system and the modeling results from $\S$\ref{section:results}. 

\begin{figure*}
        %\vspace{-1cm}
        %\hspace{-1cm}}
    \includegraphics[width=1.0\textwidth]{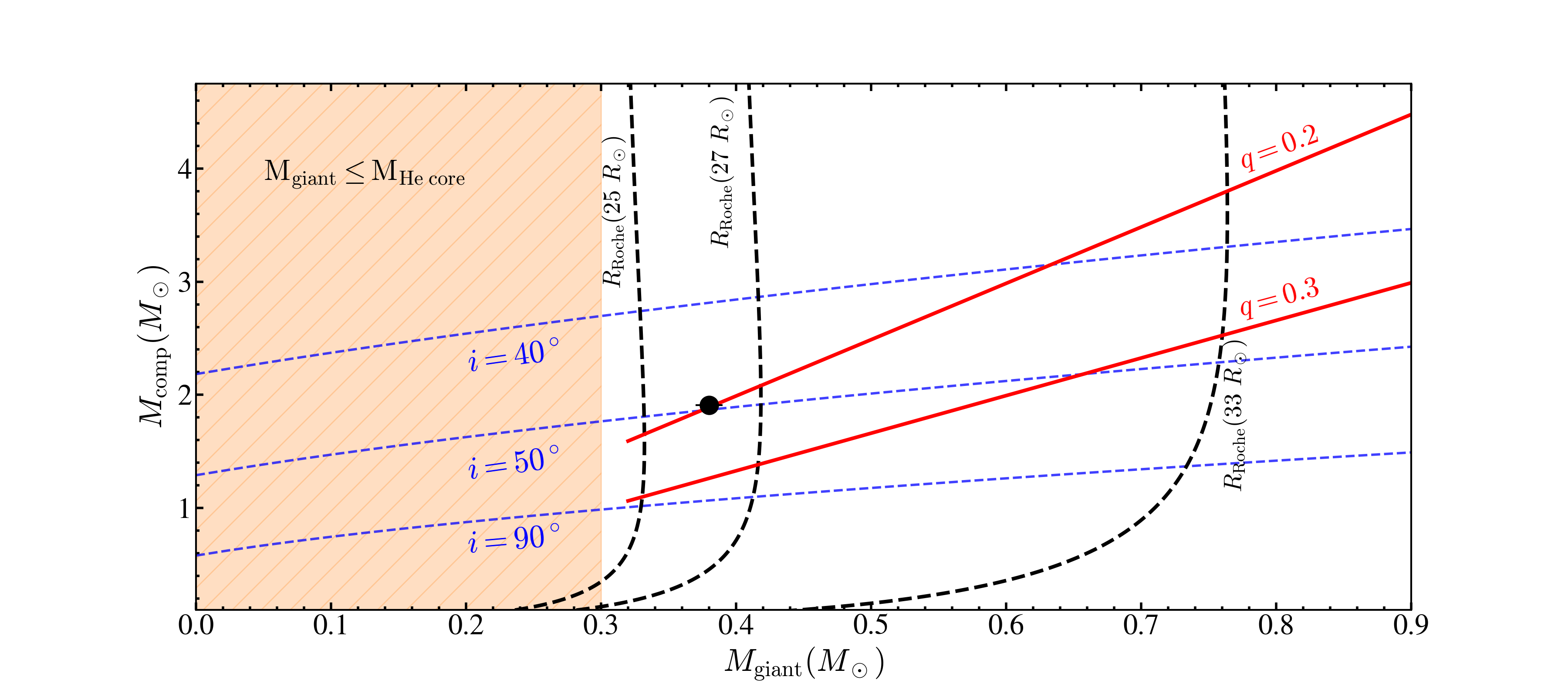}
        \vspace{-0.7cm}
    \caption{The mass of the companion ($M_{\rm comp}$) as a function of the mass of the giant ($M_{\rm giant}$). The binary mass function at inclinations of $i=90^\circ$, $50^\circ$ and $40^\circ$ are shown as the blue dashed lines. Giants with radii of $R=25~R_\odot$, $R=29~R_\odot$ or $R=33~R_\odot$ will overfill their Roche lobes of $R_{\rm Roche}(R)$ if they have masses less than the Roche limits given by the dashed curves. The companion masses for a mass ratio of $q=0.2$ and $q=0.3$ are shown as the red lines. The orange shaded region highlights giant masses that are not allowed by the condition $M_{\rm giant}\gtrsim M_{\rm core}\simeq0.32~M_\odot$ (Equation \ref{eq:coremass}). The derived values for $M_{\rm comp}$ and $M_{\rm giant}$ from the PHOEBE model ($\S$\ref{section:phoebe}) is shown in black.}
    \label{masses}
\end{figure*}

Figure \ref{masses} summarizes the overall landscape of the constraints on the masses of the giant and the companion. There are several constraints on the giant mass that are independent of single star evolution models and that we show in Figure \ref{masses}. First, the giant is on the upper RGB and not on the red clump (Figure \ref{loggteff}), so it has a degenerate core. The luminosity of a solar metallicity giant with a degenerate core of mass $M_{\rm core}$ is \citep{Boothroyd1988}, 
\begin{equation}
    \label{eq:coremass}
    \bigg(\frac{L}{L_\odot}\bigg)\approx 1170 \bigg(\frac{M_{\rm core}}{0.4~M_\odot}\bigg)^{7},
\end{equation} 
so we must have $M_{\rm giant}>M_{\rm core}\simeq0.30~M_\odot$ for $L\sim150~L_\odot$. The core mass--luminosity relationship is weakly dependent on metallicity. For example, this limit changes to $M_{\rm giant}>M_{\rm core}\simeq0.32~M_\odot$ for a metal-poor giant with $Z=0.001$. This region is denoted by the orange hatched region in Figure \ref{masses}. Using the \citet{Eggleton1983} approximation for the Roche limit we illustrate the Roche radii for 25, 27, and 33\,R$_\odot$ with the black dashed lines. We see in Figure \ref{masses} that the giant must be more massive than $M_{\rm giant} \gtorder 0.32~M_\odot$ for the radius of $R_{\rm giant}{\simeq}25~R_\odot$ from the SED ($\S$\ref{section:giant}). Note that this limit is identical to the argument that the mean density of a Roche lobe-filling star is largely determined by the period (\citealt{Eggleton1983}), which was recently used by \cite{El-Badry2021} in arguing against the NGC~1850 BH1 candidate discussed in the introduction.  For $q\simeq0.2$ (Figure \ref{rvsb2}), the giant's mass is $M_{\rm giant}\simeq0.4~M_\odot$ and the companion has $M_{\rm comp}\simeq1.9~M_\odot$ for $i\simeq50^\circ$. This model is shown as the large filled black circle in Figure \ref{masses}.

That the companion RV signal in Figure \ref{rvsb2} corresponds to a simple reflex motion of the giant and the associated absorption features are narrow rules out the possibility that the companion is a binary. If the companion was itself a binary, the radial velocities would then be dominated by the motion of this inner binary and so we would not observe what seems to be the reflex motion of the outer binary. For example, an equal mass binary consisting of two $1.3~M_\odot$ stars with a separation $a_{\rm comp}=30~R_\odot$ will have an orbital period of ${\sim}11.8$~days. The velocity of each component in this binary is ${\sim}64\, \rm km \,s^{-1}$ (${\sim}49\, \rm km \,s^{-1}$) for $i=90^\circ$ ($i=50^\circ$), which exceeds the reflex motion (${\sim}8\, \rm km \,s^{-1}$) and the broadening ($v_{\rm rot} \, \textrm{sin} \, i{\sim}18 \, \rm km \,s^{-1}$) observed in the secondary absorption lines. All other configurations of equal mass inner binaries will have shorter orbits and therefore, larger velocities.

While it is unlikely in the case of 2M0412, we investigated whether the second set of absorption lines could emerge from an accretion disk/flow surrounding a compact object. In $\S$\ref{section:balmer}, we show that the source has a broad, double-peaked $\rm H\alpha$ emission line in our spectra, and argue that it likely emerges due to an accretion disk. While absorption lines are rarely seen in the spectra of accretion disks, it is not unprecedented. For example, \citet{Strader2016} identified two sets of absorption lines in spectra of the neutron star X-ray binary 3FGL J0427.9$-$6704, which corresponded to the stellar component and the accretion disk. The second set of absorption lines track the reflex motion of the neutron star primary and also matched the velocities derived from the emission features associated with the accretion disk. Given that we are able to model the absorption lines with that of a stellar spectrum, it is unlikely that they emerge from an accretion flow.

\begin{figure*}
	\includegraphics[width=0.98\textwidth]{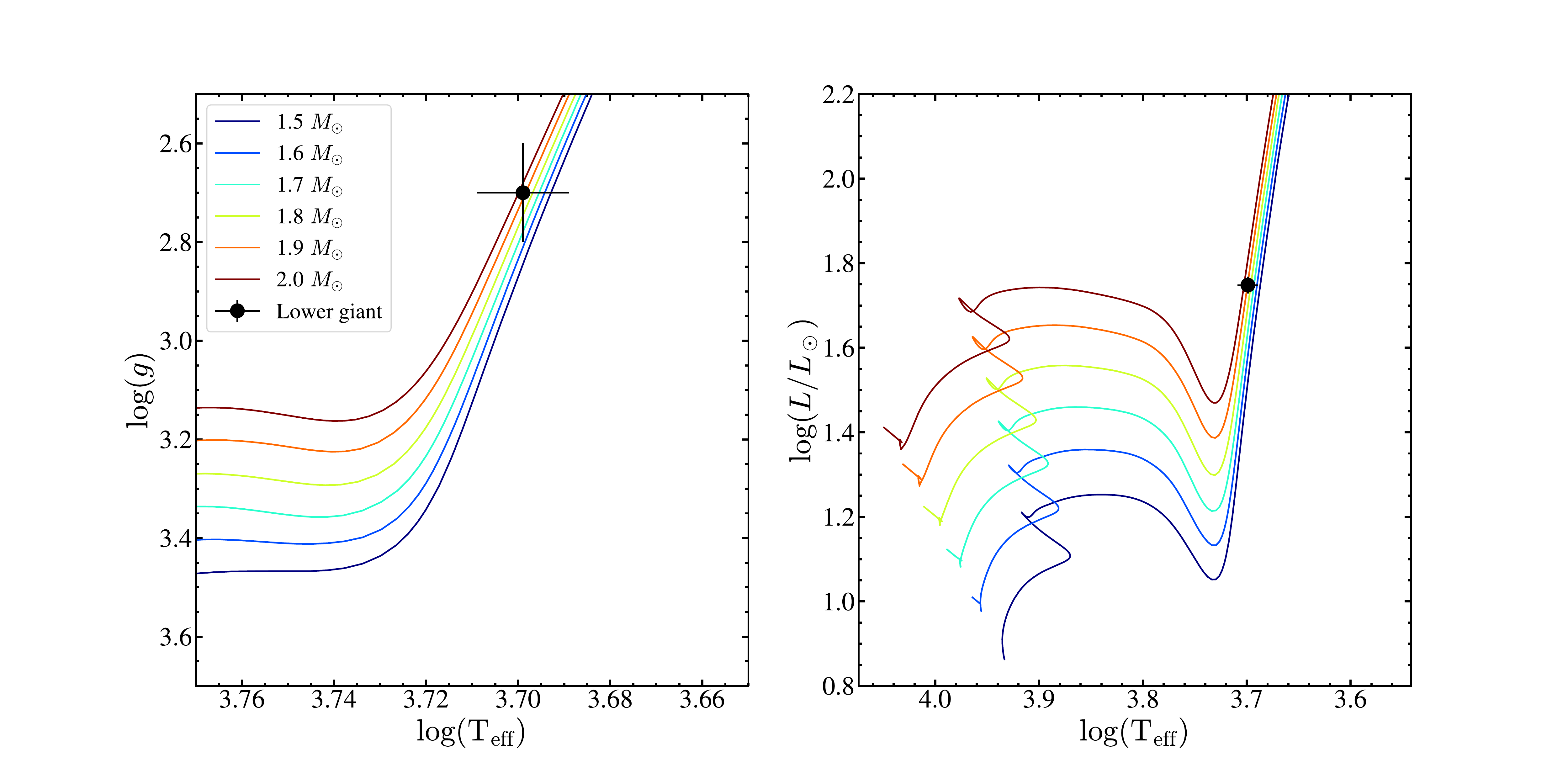}
    \caption{\textit{Left:} $\log(g)$ vs. $T_{\rm eff}$ and, \textit{Right:} $\log(L)$ vs. $\log(T_{\rm eff})$. The position of the companion to 2M0412 is shown in black. MIST \citep{Dotter2016,Choi2016} stellar tracks for $\rm [Fe/H]=-0.5$ are shown.}
    \label{loggteffsg}
\end{figure*}

%SComparison to stellar models?
The properties of the companion to 2M0412 are well explained by single star MIST stellar models. Figure \ref{loggteffsg} shows the position of the companion in $\log(g)$ vs. $T_{\rm eff}$ and $\log(L)$ vs. $T_{\rm eff}$ compared to the MIST \citep{Dotter2016,Choi2016}) stellar tracks at $\rm [Fe/H]=-0.5$. Its position is consistent with that of a lower giant with $M_{\rm comp}\gtrsim1.9~M_\odot$. However, we note that the companion appears to rotate slower ($v_{\rm rot}\sin(i)\simeq18$~km/s) than expected for an object that has accreted significant mass. \citet{ElBadry2022} presents a detailed comparison between the properties of the companion and stellar models, and interprets the slow rotation of the companion as a result of tidal interactions.

\section{Conclusions}
\label{section:summary}

The evolved, stripped red giant 2M04123153+6738486 with $T_{\rm eff, giant}\simeq4000$~K, $L_{\rm giant}\simeq150~L_\odot$, $R_{\rm giant}\simeq25~R_\odot$ is in a $P=81.2$~d interacting binary with a lower giant companion ($T_{\rm eff, giant}\simeq5000$~K, $L_{\rm giant}\simeq60~L_\odot$, $R_{\rm comp}\simeq11~R_\odot$). The ASAS-SN, ATLAS, \textit{TESS} and ZTF light curves show that 2M0412 is a Roche Lobe filling ellipsoidal variable ($\S$\ref{section:phoebe}, Figure \ref{lcs}). The high-resolution Keck/HIRES and LBT/PEPSI spectra indicate that the giant is rapidly rotating ($\S$\ref{section:phoebe}). 

We also detect a secondary component in the spectra (Figure \ref{rvsb2}) and the RV cross-correlation function (Figure \ref{rvccf}), which implies a mass ratio of $q\simeq0.201\pm0.010$ (Figure \ref{rvsb2}). Given that this second component traces the reflex motion expected for a simple binary, it is very unlikely that the companion can itself be a binary.  

\verb"PHOEBE" models of the ellipsoidal variability ($\S$\ref{section:phoebe}, give an orbital inclination $i_{\rm orb}\simeq49.4^\circ$, the radius of the giant $R_{\rm giant}\simeq26~R_\odot$, the mass of the giant $M_{\rm giant}\simeq0.38~M_\odot$ and a companion mass of $M_{\rm comp}=1.9~M_\odot$ for $q\simeq0.2$.  The inferred mass of the red giant ($M_{\rm giant}\simeq0.38~M_\odot$) is far less than expected based on its luminosity and single star evolutionary tracks which implies that the giant has been heavily stripped by binary interaction. 

We can find BPASS binary models very similar to the observed system with both stellar and compact object companions, and in all cases the giant has been partly stripped by mass transfer. MESA models of the cases with a stellar companion, which are needed to properly predict the luminosity and temperature of the companion, predict that a stellar companion would be significantly hotter ($\log T_{\rm eff}\gtrsim4.0$) and more luminous ($\log (L/L_\odot)\gtrsim1.5$) than allowed by the observed SED. A more detailed exploration of parameter space can, however, find binaries that match our observations \citep{ElBadry2022}.

We also find a broad, phase variable $\rm H\alpha$ emission line ($\S$\ref{section:balmer}, Figure \ref{spechalpha}) which is likely due to an accretion disk. The properties of the $\rm H\alpha$ emission are consistent with Roche lobe overflow from the giant. We derive a \textit{Swift} XRT X-ray flux limit corresponding to $L_{\rm X}\lesssim2.9\times10^{31}~\rm ergs~s^{-1}$ ($\S$\ref{section:xrays}). 

\section*{Acknowledgements}

We thank the anonymous referee for their comments. We thank Dr. J. Strader for a careful reading of this manuscript. We thank Dr. J. J. Eldridge for useful discussions on the BPASS models. We thank Dr. Marc Pinsonneault and Lyra Cao for useful discussions. We thank Las Cumbres Observatory and its staff for their continued support of ASAS-SN. ASAS-SN is funded in part by the Gordon and Betty Moore Foundation through grants GBMF5490 and GBMF10501 to the Ohio State University, and also funded in part by the Alfred P. Sloan Foundation grant G-2021-14192.

TJ, KZS and CSK are supported by NSF grants AST-1814440 and 
AST-1908570. TJ acknowledges support from the Ohio State Presidential Fellowship. TAT is supported in part by NASA grant 80NSSC20K0531. TAT acknowledges previous support from Scialog Scholar grant 24216 from the Research Corporation, from which this effort germinated.  J.T.H. is supported by NASA award 80NSSC21K0136.
D.H. acknowledges support from the Alfred P. Sloan Foundation, the National Aeronautics and Space Administration (80NSSC18K1585, 80NSSC19K0379), and the National Science Foundation (AST-1717000). CB acknowledges support from the National Science Foundation grant AST-1909022.

Parts of this research were supported by the Australian Research Council Centre of Excellence for All Sky Astrophysics in 3 Dimensions (ASTRO 3D), through project number CE170100013 and the Australian Research Council Centre of Excellence for Gravitational Wave Discovery (OzGrav), through project number CE170100004.

Funding for the Sloan Digital Sky Survey IV has been provided by the Alfred P. Sloan Foundation, the U.S. Department of Energy Office of Science, and the Participating Institutions. SDSS-IV acknowledges
support and resources from the Center for High-Performance Computing at
the University of Utah. The SDSS web site is www.sdss.org.

SDSS-IV is managed by the Astrophysical Research Consortium for the 
Participating Institutions of the SDSS Collaboration including the 
Brazilian Participation Group, the Carnegie Institution for Science, 
Carnegie Mellon University, the Chilean Participation Group, the French Participation Group, Harvard-Smithsonian Center for Astrophysics, 
Instituto de Astrof\'isica de Canarias, The Johns Hopkins University, Kavli Institute for the Physics and Mathematics of the Universe (IPMU) / 
University of Tokyo, the Korean Participation Group, Lawrence Berkeley National Laboratory, 
Leibniz Institut f\"ur Astrophysik Potsdam (AIP),  
Max-Planck-Institut f\"ur Astronomie (MPIA Heidelberg), 
Max-Planck-Institut f\"ur Astrophysik (MPA Garching), 
Max-Planck-Institut f\"ur Extraterrestrische Physik (MPE), 
National Astronomical Observatories of China, New Mexico State University, 
New York University, University of Notre Dame, 
Observat\'ario Nacional / MCTI, The Ohio State University, 
Pennsylvania State University, Shanghai Astronomical Observatory, 
United Kingdom Participation Group,
Universidad Nacional Aut\'onoma de M\'exico, University of Arizona, 
University of Colorado Boulder, University of Oxford, University of Portsmouth, 
University of Utah, University of Virginia, University of Washington, University of Wisconsin, 
Vanderbilt University, and Yale University.

This work has made use of data from the European Space Agency (ESA)
mission {\it Gaia} (\url{https://www.cosmos.esa.int/gaia}), processed by
the {\it Gaia} Data Processing and Analysis Consortium. This publication makes 
use of data products from the Two Micron All Sky Survey, as well as
data products from the Wide-field Infrared Survey Explorer.
This research was also made possible through the use of the AAVSO Photometric 
All-Sky Survey (APASS), funded by the Robert Martin Ayers Sciences Fund. 

The LBT is an international collaboration among institutions in the
United States, Italy and Germany. LBT Corporation partners are: The
University of Arizona on behalf of the Arizona Board of Regents;
Istituto Nazionale di Astrofisica, Italy; LBT Beteiligungsgesellschaft,
Germany, representing the Max-Planck Society, The Leibniz Institute for
Astrophysics Potsdam, and Heidelberg University; The Ohio State
University, representing OSU, University of Notre Dame, University of
Minnesota and University of Virginia.

PEPSI was made possible by funding through the State of
Brandenburg (MWFK) and the German Federal Ministry of Education and
Research (BMBF) through their Verbundforschung grants 05AL2BA1/3 and
05A08BAC.

This research is based on observations made with the \textit{Neil Gehrels Swift Observatory}, obtained from the MAST data archive at the Space Telescope Science Institute, which is operated by the Association of Universities for Research in Astronomy, Inc., under NASA contract NAS 5-26555. This paper includes data collected with the \textit{TESS} mission, obtained from the MAST data archive at the Space Telescope Science Institute (STScI). Funding for the TESS mission is provided by the NASA Explorer Program. STScI is operated by the Association of Universities for Research in Astronomy, Inc., under NASA contract NAS 5-26555.

Some of the data presented herein were obtained at the W. M. Keck Observatory, which is operated as a scientific partnership among the California Institute of Technology, the University of California and the National Aeronautics and Space Administration. The Observatory was made possible by the generous financial support of the W. M. Keck Foundation.

Based on observations obtained with the Samuel Oschin Telescope 48-inch and the 60-inch Telescope at the Palomar Observatory as part of the Zwicky Transient Facility project. ZTF is supported by the National Science Foundation under Grant No. AST-2034437 and a collaboration including Caltech, IPAC, the Weizmann Institute for Science, the Oskar Klein Center at Stockholm University, the University of Maryland, Deutsches Elektronen-Synchrotron and Humboldt University, the TANGO Consortium of Taiwan, the University of Wisconsin at Milwaukee, Trinity College Dublin, Lawrence Livermore National Laboratories, and IN2P3, France. Operations are conducted by COO, IPAC, and UW.

The authors wish to recognize and acknowledge the very significant cultural role and reverence that the summit of Maunakea has always had within the indigenous Hawaiian community.  We are most fortunate to have the opportunity to conduct observations from this mountain.

We thank the ZTF and ATLAS projects for making their light curve data publicly available.
This research has made use of the VizieR catalogue access tool, CDS, Strasbourg, France. 
This research also made use of Astropy, a community-developed core Python package for 
Astronomy \citep{astropy:2013,astropy:2018}.

%%%%%%%%%%%%%%%%%%%%%%%%%%%%%%%%%%%%%%%%%%%%%%%%%%
\section*{Data Availability}

The photometric data are all publicly available. The spectra will be shared on reasonable request to the corresponding author.

%%%%%%%%%%%%%%%%%%%% REFERENCES %%%%%%%%%%%%%%%%%%

% The best way to enter references is to use BibTeX:

\bibliographystyle{mnras}
\bibliography{refbh} % if your bibtex file is called example.bib

% Alternatively you could enter them by hand, like this:
% This method is tedious and prone to error if you have lots of references
%\begin{thebibliography}{99}
%\bibitem[\protect\citeauthoryear{Author}{2012}]{Author2012}
%Author A.~N., 2013, Journal of Improbable Astronomy, 1, 1
%\bibitem[\protect\citeauthoryear{Others}{2013}]{Others2013}
%Others S., 2012, Journal of Interesting Stuff, 17, 198
%\end{thebibliography}

%%%%%%%%%%%%%%%%%%%%%%%%%%%%%%%%%%%%%%%%%%%%%%%%%%

%%%%%%%%%%%%%%%%% APPENDICES %%%%%%%%%%%%%%%u%%%%%%

%%%%%%%%%%%%%%%%%%%%%%%%%%%%%%%%%%%%%%%%%%%%%%%%%%

% Don't change these lines
\bsp	% typesetting comment
\label{lastpage}
\end{document}